\documentclass[aps,10pt,prd,letterpaper,preprintnumbers,amsmath,amssymb,floatfix,nofootinbib]{revtex4}
\usepackage{epsfig}
\usepackage{amsfonts}
\usepackage{amsmath}
\usepackage{amssymb}
\usepackage{slashed}
\usepackage{color,hyperref}
\hypersetup{colorlinks,citecolor= nicegreen,linkcolor= nicered}
\definecolor{nicered}{rgb}{0.7,0.1,0.1}
\definecolor{nicegreen}{rgb}{0.1,0.5,0.1}

\newcommand{\dslash}[1]{#1 \llap{/\kern-0.5pt}}
\newcommand{\Dslash}[1]{#1 \llap{/\kern+1.2pt}}
\newcommand{\DDslash}[1]{#1 \llap{/\kern+2.3pt}}
\newcommand{\dslashh}[1]{#1 \llap{/\kern+1pt}}

\newcommand{\be}{\begin{equation}}
\newcommand{\ee}{\end{equation}}
\newcommand{\bee}{\begin{equation*}}
\newcommand{\eee}{\end{equation*}}
\newcommand{\bea}{\begin{eqnarray}}
\newcommand{\eea}{\end{eqnarray}}
\newcommand{\bean}{\begin{eqnarray*}}
\newcommand{\eean}{\end{eqnarray*}}

\begin{document}

\preprint{ACFI-T14-06}
\preprint{CALT-68-2881}

\title{CPV Phenomenology of Flavor Conserving Two Higgs Doublet Models}

\author{Satoru Inoue$^{1}$}
\author{Michael J. Ramsey-Musolf$^{1,2}$}
\author{Yue Zhang$^2$}

\affiliation{\hspace{5mm}
\\
$^{1}${\it Amherst Center for Fundamental Interactions\\
Department of Physics, University of Massachusetts Amherst\\
Amherst, MA 01003 USA}\\
\\
$^{2}${\it California Institute of Technology\\
Pasadena, CA 91125 USA}\\
}

\date{\today}

\begin{abstract}
We analyze the constraints on CP-violating, flavor conserving Two Higgs Doublet Models (2HDMs) implied by measurements of Higgs boson properties
at the Large Hadron Collider (LHC) and by the non-observation of permanent electric dipole moments (EDMs) of molecules, atoms and the neutron. We find that the LHC and EDM constraints are largely complementary, with the LHC studies constraining the mixing between the neutral CP-even states and EDMs probing the effect of mixing between the CP-even and CP-odd scalars. The presently most stringent constraints are implied by the non-observation of the ThO molecule EDM signal. Future improvements in the sensitivity of neutron and diamagnetic atom EDM searches could yield competitive or even more severe constraints. We analyze the quantitative impact of hadronic and nuclear theory uncertainties on the interpretation of the latter systems and conclude that these uncertainties cloud the impact of projected improvements in the corresponding experimental sensitivities.

\end{abstract}

\maketitle

\section{Introduction}
\label{sec:intro}
With the discovery of a 125 GeV boson at the Large Hadron Collider, exploration of the dynamics of electroweak symmetry-breaking (EWSB) is front and center in particle physics. Although the properties of the new boson thus far agree with expectations for the Standard Model Higgs boson, it is possible that it constitutes but one state of a richer scalar sector. Perhaps, the most widely-studied extended scalar sector is the two-Higgs doublet model (2HDM). The collider and low-energy phenomenology of the 2HDM have been extensively analyzed over the years~\cite{hunter, Branco:2011iw}, while one version of this paradigm appears in an equally important scenario for physics beyond the Standard Model (BSM), the minimal supersymmetric Standard Model. 

One of the more interesting features of the 2HDM is the presence of new sources of CP-violation (CPV) beyond that of the Standard Model Cabibbo-Kobayashi-Maskawa (CKM) matrix and the QCD \lq\lq $\theta$-term". It is well-known that BSM CPV is required to account for the observed excess of matter over anti-matter in the present Universe. If realized in nature, the 2HDM may provide the necessary CPV and enable the generation of the matter-antimatter asymmetry during the era of EWSB~\cite{Morrissey:2012db, Shu:2013uua, Fromme:2006cm}, and possibly the co-generation of both baryonic and dark matter in the universe~\cite{Cheung:2013dca}. If so, then the 2HDM CPV may have observable signatures in laboratory tests. At the energy frontier, CPV correlations associated with the production and decay of the lightest neutral scalar may be accessible at the LHC and/or a future high intensity $e^+e^-$ collider. At the low-energy \lq\lq intensity frontier", searches for the permanent electric dipole moments (EDMs) of atoms, molecules and nucleons provide a powerful indirect probe~\cite{Engel:2013lsa, Fukuyama:2012np, Pospelov:2005pr, Bernreuther:1990jx}. Indeed, EDM searches are entering a new era of sensitivity, with the recent report by the ACME collaboration of a ten times tighter limit on the electron EDM~\cite{Baron:2013eja}\footnote{The experiment actually constrains the  ThO molecule response to an external electric field . In general, the ThO response is dominated by two operators including the electron EDM and an electron-quark interaction. For the 2HDM, the electron EDM gives by far the larger contribution.}, representing a harbinger of even more powerful probes in the future. Searches for the permanent EDM of the neutron are underway at a variety of laboratories around the world, with goals of one-to-two orders of magnitude improvements in sensitivity. Similarly, experiments are underway to carry out improved and/or new searches for the EDMs of Mercury, Xenon, and Radium,  while longer term efforts to develop storage ring probes of the proton and light nuclei EDMs are being pursued~\cite{Kumar:2013qya}. 

With this context in mind, it is timely to investigate the present and prospective probes of CPV in the 2HDM. In what follows, we report on such a study, focusing on the implications of present and prospective EDM searches while taking into account the LHC constraints on the properties of the 125 GeV boson. For concreteness, we consider a $Z_2$-symmetric variant of the 2HDM that evades potentially problematic flavor changing neutral currents (FCNCs) but still allows for new CPV associated with the scalar potential, accommodates the present 125 GeV boson properties, and retains a rich phenomenology for future studies. 

Under these assumptions, we explore scenarios wherein there exists only one physical CPV phase associated with the scalar potential. The resulting scalar spectrum contains three neutral states that are CPV mixtures of neutral scalar and pseudo-scalars and one pair of charged scalars. The scalar sector is then characterized by nine-independent parameters that can be related to the parameters in the potential using the conditions for EWSB. We take these parameters to be the four scalar masses; the mixing angle $\alpha_b$ that governs the CP-odd admixture of the 125 GeV scalar; the combined vacuum expectation value (vev) of the two neutral scalars, $v=246$ GeV; the conventional 2HDM mixing angles $\beta$ and $\alpha$; and a parameter $\nu$ (also defined below) that characterizes the degree to which the heavier states decouple from the low-energy effective theory, leaving 125 GeV boson as the only accessible state. Note that $\alpha_b\to 0$ when $\delta\to 0$ and that under our choice of the independent parameters, $\alpha_b$ encodes the effects of the sole physical phase in the scalar potential. 

From our analysis, we find that
\begin{itemize}
\item Fits to the properties of the observed 125 GeV boson generally favor scenarios in which $\alpha\approx\beta-\pi/2$
\item For fixed values of the scalar masses, null results for EDMs yield constraints in the $\sin\alpha_b$--$\tan\beta$ plane.  
\item The assumption that loops involving the lightest neutral scalar are dominant over those involving the remaining 2HDM scalar states does not hold in general. In the electron EDM case, for example, loops involving the heavier states may yield the largest contribution for moderate-to-large $\tan\beta$. In short, the \lq\lq light Higgs effective theory" is not necessarily effective in this context. 
\item At present the ThO result yields the strongest constraints on the CPV parameter space. Future EDM searches could significantly extend this reach, particularly for the type II 2HDM. An order-of-magnitude improvement in the sensitivity of the neutron and $^{199}$Hg EDM searches would probe regions presently allowed by the ThO limit. A factor of 100 more sensitive neutron EDM search would go well beyond the present constraints, and in the event of a null result, would restrict $|\sin\alpha_b|$ to less than a few $\times 10^{-3}$. Successful completion of the Argonne $^{225}$Ra EDM search at its design sensitivity would reach well beyond the ACME constraints as well as the possible ten times better neutron and $^{199}$Hg search for the type II case, though a 100 times more sensitive neutron EDM search would surpass the radium reach. For the type I model, the ACME constraints would survive even the future neutron, mercury, and radium experiments. Thus, a non-zero result for any of the latter searches would indicate the presence of a type II rather than a type I 2HDM.
\item Determination of the diamagnetic atom ($^{199}$Hg, $^{225}$Ra, {\em etc}) and neutron EDM sensitivities is subject to considerable hadronic and nuclear many-body uncertainties. Those associated with the interpretation of the paramagnetic system (ThO) results are less significant. Consequently, the aforementioned statements about the relative sensitivities of future searches are provisional. Definitive conclusions will require substantial improvements in hadronic and nuclear many-body computations.
\end{itemize}  

We organize the discussion of the analysis leading to these observations as follows. In Section \ref{sec:model} we analyze the general features of the $Z_2$-symmetric 2HDM, including the constraints of the EWSB conditions, the choice of independent parameters, and the structure of the interactions. Section \ref{sec:obs} gives the relationship of the independent parameters and interactions to the observables of interest, including the Higgs boson event rates at the LHC (\ref{sec:lhc}), the low-scale effective operators that ultimately induce EDMs of various light quark and lepton systems and their renormalization group evolution (\ref{running}), the sensitivity of these systems to the effective operators and the corresponding EDM constraints (\ref{sec:edmconstraints}). Section \ref{sec:results} gives the resulting constraints on the relevant parameter space. In particular we call the reader's attention to Figures \ref{bound&uncertainty} and \ref{Fig:Comb}. The former gives the present constraints from ThO, the neutron, and $^{199}$Hg in the $\sin\alpha_b$--$\tan\beta$ plane, including the hadronic and nuclear theory uncertainties. The latter shows the prospective impact of future neutron, $^{199}$Hg, and $^{225}$Ra searches in comparison with the present ThO constraints. In this section, we also present an \lq\lq anatomy" of the electron, neutron and diamagnetic EDMs in terms of the 2HDM degrees of freedom as well as the various low-energy effective operators. We summarize our conclusions in Section \ref{sec:summary}. Expressions for the effective operator Wilson coefficients are given in an Appendix.





\section{2HDM Framework}
\label{sec:model}

\subsection{Scalar potential}
In this work, we consider the flavor-conserving 2HDM
in order to avoid problematic flavor-changing neutral currents (FCNCs). As observed by Glashow and Weinberg (GW)~\cite{glashow}, one may avoid tree-level FCNCs if diagonalization of the fermion mass matrices leads to flavor diagonal Yukawa interactions. One approach\footnote{Another approach is to have 2HDM at the electroweak scale without the $Z_2$ symmetry is to assume minimal flavor violation, flavor alignment or other variants. We do not discuss this possibility, but refer to~\cite{Buras:2010zm, Cline:2011mm, Jung:2013hka} for recent phenomenological studies.} to realizing this requirement is to impose a $Z_2$ symmetry on the scalar potential together with an appropriate extension to the Yukawa interactions (see below). In this scenario, however, one obtains no sources of CPV beyond the SM CKM complex phase. Consequently, we introduce a soft $Z_2$-breaking term that yields non-vanishing CPV terms in the scalar sector~\cite{Accomando:2006ga}. 

To that end, we choose a scalar field basis in which the two Higgs doublets $\phi_{1,2}$ are oppositely charged under the the $Z_2$ symmetry:
\be
\label{eq:Z2a}
\phi_1\to - \phi_1\qquad \mathrm{and}\qquad \phi_2\to\phi_2\ \ \ ,
\ee
though this symmetry will in general have a different expression in another basis obtained by the transformation $\phi_j= U_{jk} \phi_k^\prime$. For example, taking
\be
U = \frac{1}{\sqrt{2}} \left(
\begin{array}{cc}
-1 & 1\\ 
1 & 1 
\end{array}\right)  \ ,
\label{eq:transform}
\ee
the transformation (\ref{eq:Z2a}) corresponds to
\be
\label{eq:Z2b}
\phi_1^\prime\leftrightarrow\phi_2^\prime\ \ \ .
\ee

We then take the Higgs potential to have the form
\begin{eqnarray}                    \label{Eq:gko-pot}
V&=&\frac{\lambda_1}{2}(\phi_1^\dagger\phi_1)^2
+\frac{\lambda_2}{2}(\phi_2^\dagger\phi_2)^2
+\lambda_3(\phi_1^\dagger\phi_1) (\phi_2^\dagger\phi_2) 
+\lambda_4(\phi_1^\dagger\phi_2) (\phi_2^\dagger\phi_1) +\frac{1}{2}\left[\lambda_5(\phi_1^\dagger\phi_2)^2+{\rm h.c.}\right] \nonumber \\
&&
-\frac{1}{2}\left\{m_{11}^2(\phi_1^\dagger\phi_1)
+\left[m_{12}^2 (\phi_1^\dagger\phi_2)+{\rm h.c.}\right]
+m_{22}^2(\phi_2^\dagger\phi_2)\right\}. 
\end{eqnarray}
The complex coefficients in the potential are $m_{12}^2$ and $\lambda_5$.  
In general, the presence of the $\phi_1^\dag \phi_2$ term, in conjunction with the $Z_2$-conserving quartic interactions, will induce
other $Z_2$-breaking quartic operators at one-loop order. Simple power counting implies that the responding coefficients are finite with magnitude proportional to 
$m_{12}^2 \lambda_k/(16\pi^2)$. Given the $1/16\pi^2$ suppression, we will restrict our attention to the tree-level $Z_2$-breaking bilinear term. 

It is instructive to identify the CPV complex phases that are invariant under a rephasing of the scalar fields. To that end, we perform an SU(2$)_L\times$U(1$)_Y$ transformation to a basis where the vacuum expectation value (vev) of the neutral component of $\phi_1$ is real while that associated with the neutral component of $\phi_2$ is in general complex:
\begin{eqnarray}\label{decompose}
\phi_1=\begin{pmatrix}
H_1^+ \\
\frac{1}{\sqrt2} (v_1 + H_1^0 + i A_1^0)
\end{pmatrix}, \ \ 
\phi_2=\begin{pmatrix}
H_2^+ \\
\frac{1}{\sqrt2} (v_2 + H_2^0 + i A_2^0)
\end{pmatrix} \ ,
\end{eqnarray}
where $v=\sqrt{|v_1|^2 +|v_2|^2} = 246\,$GeV, $v_1=v_1^\ast$ and $v_2=|v_2| e^{i\xi}$. It is apparent that in general $\xi$ denotes the relative phase of $v_2$ and $v_1$. Under the global rephasing transformation
\be
\label{eq:rephase}
\phi_j = e^{i \theta_j}\, \phi_j^\prime \ ,
\ee
the couplings $m_{12}^2$ and $\lambda_5$ can be redefined to absorb the global phases
\be
(m_{12}^2)^\prime = e^{i(\theta_2-\theta_1)} m_{12}^2, \ \ \ \ \  \lambda_5^\prime = e^{2i(\theta_2-\theta_1)}\lambda_5 \ ,
\ee
so that the form of the potential is unchanged. It is then straightforward to observe that there exist two rephasing invariant complex phases:
\bea
\nonumber
\delta_1 & = & \mathrm{Arg}\left[\lambda_5^\ast(m_{12}^2)^2\right] \ ,\\
\delta_2 & = & \mathrm{Arg}\left[\lambda_5^\ast(m_{12}^2) v_1 v_2^\ast\right]\ .
\eea
For future purposes, we emphasize that the value of $\xi$ is not invariant. 

Denoting $\tan\beta= |v_2|/|v_1|$, 
the minimization conditions in the $H^0_k$ and $A^0_k$ directions give us the relations
\begin{eqnarray}\label{mini}
&&m_{11}^2 = \lambda_1 v^2 \cos^2\beta + (\lambda_3 + \lambda_4) v^2 \sin^2\beta - {\rm Re} (m_{12}^2 e^{i\xi}) \tan\beta + {\rm Re} (\lambda_5 e^{2i\xi}) v^2\sin^2\beta \ ,
\label{mini1}\\
&&m_{22}^2 = \lambda_2 v^2 \sin^2\beta + (\lambda_3 + \lambda_4) v^2 \cos^2\beta - {\rm Re} (m_{12}^2 e^{i\xi}) \cot\beta + {\rm Re} (\lambda_5 e^{2i\xi}) v^2\cos^2\beta \ ,
\label{mini2}\\
&&{\rm Im} (m_{12}^2 e^{i\xi})=v^2 \sin\beta\cos\beta {\rm Im} ( \lambda_5 e^{2i\xi} ) \ . \label{mini3} 
\end{eqnarray}
From the last equation, it is clear that the phase $\xi$ can be solved for given the complex parameters $m_{12}^2$ and $\lambda_5$. It is useful, however, to express this condition in terms of the $\delta_k$:
\be
\label{eq:invar1}
|m_{12}^2| \sin(\delta_2-\delta_1) = |\lambda_5 v_1 v_2|\sin(2\delta_2-\delta_1)\ \ \ .
\ee
In the limit that the $\delta_k$ are small but non-vanishing that will be appropriate for our later phenomenological discussion, Eq.~(\ref{eq:invar1}) then implies 
\be
\label{eq:invar2}
\delta_2 \approx \frac{1-\left| \frac{\lambda_5 v_1 v_2}{m_{12}^2} \right|}{1-2 \left| \frac{\lambda_5 v_1 v_2}{m_{12}^2} \right|} \delta_1\ \ \ ,
\ee
so that there exists only one independent CPV phase in the theory after EWSB.

A special case arises when $\delta_1=0$. In this case, Eq.~(\ref{eq:invar1}) implies that 
\begin{eqnarray}
|m_{12}^2| \sin(\delta_2)=|\lambda_5 v_1 v_2| \sin(2\delta_2) \ ,
\end{eqnarray}
or
\begin{eqnarray}
\label{eq:scpv1}
\cos \delta_2 =\frac{1}{2}\left| \frac{m_{12}^2}{\lambda_5 v_1 v_2} \right| \ .
\end{eqnarray}
When the right-hand side is less than 1, $\delta_2$ has solutions two solutions of equal magnitude and opposite sign, corresponding to the presence of spontaneous CPV (SCPV)~\cite{Lee:1974jb, Grzadkowski:2013rza}:
\begin{eqnarray}
\label{eq:scpv2}
 \delta_2 =\pm\arccos \left(\frac{1}{2}\left| \frac{m_{12}^2}{\lambda_5 v_1 v_2} \right|\right) = \pm 
 \left(\frac{1}{2}\left| \frac{m_{12}^2}{\lambda_5 v^2\cos\beta\sin\beta} \right|\right)\ \ \ .
\end{eqnarray}
To the extent that the vacua associated with the two opposite sign solutions are degenerate, one would expect the existence of cosmological domains~\cite{Zeldovich:1974uw} associated with these two vacua. Persistence of the corresponding domain walls to late cosmic times is inconsistent with the observed homogeneity of structure and isotropy of the cosmic microwave background. Consequently, parameter choices leading to $\delta_1=0$ but $\delta_2\not=0$ should be avoided. In practice, we will scan over model parameters when analyzing the EDM and LHC constraints. As a check, we have performed a scan with $10^6$ points and find less than ten that give $\delta_1=0$. We are, thus, confident that the general features of our phenomenological analysis are consistent with the absence of problematic SCPV domains.

Henceforth, for simplicity, we utilize the rephasing invariance of the $\delta_k$ and work in a basis where $\xi=0$. In this basis, the phases of $m_{12}^2$ and $\lambda_5$ are redefined and related by Eq.~(\ref{mini3}). As we discuss below, we will trade the resulting dependence of observables on $\delta_1$ [and $\delta_2$ {\em via} $\delta_1$ in Eq.~(\ref{eq:invar2})] for one independent angle in the transformation that diagonalizes the neutral scalar mass matrix.


\subsection{Scalar spectrum}

After EWSB, the diagonalization of the $2\times2$ charged Higgs mass matrix yields the physical charged scalar and Goldstone modes,
\begin{eqnarray}
H^+ = -\sin\beta H_1^+ + \cos\beta H_2^+, \ \ \ G^+ = \cos\beta H_1^+ + \sin\beta H_2^+ \ ,
\end{eqnarray}
The charged scalar has a mass
\begin{eqnarray}\label{MC}
m^2_{H^+} = \frac{1}{2} \left(2\nu - \lambda_4 - {\rm Re}\lambda_5 \right)  v^2, \ \ \ \nu \equiv\frac{{\rm Re}m_{12}^2 \csc\beta \sec\beta}{{2v^2}} \ .
\end{eqnarray}

For the neutral Higgs sector, the mixing between CP odd components yields the Goldstone $G^0$ and an orthogonal combination $A^0$, where
\begin{eqnarray}
A^0 = -\sin\beta A_1^0 + \cos\beta A_2^0, \ \ \ G^0 = \cos\beta A_1^0 + \sin\beta A_2^0 \ .
\end{eqnarray}
In the presence of explicit CP violation, $A^0$ is not yet a mass eigenstate. It will further mix with the CP even eigenstates $H_1^0$, $H_2^0$.
The $3 \times 3$ neutral mass matrix in the basis of $\{H_1^0, H_2^0, A^0\}$ is
\begin{equation}\label{MM}
{\cal M}^2=v^2
\begin{pmatrix}
\lambda_1c_\beta^2+\nu s_\beta^2 & (\lambda_{345} - \nu)c_\beta s_\beta & -\frac{1}{2}{\rm Im}\lambda_5 \, s_\beta \\
(\lambda_{345} - \nu)c_\beta s_\beta & \lambda_2 s_\beta^2+\nu c_\beta^2 & -\frac{1}{2}{\rm Im}\lambda_5\, c_\beta\\
-\frac{1}{2}{\rm Im}\lambda_5\, s_\beta & -\frac{1}{2}{\rm Im}\lambda_5\, c_\beta & -{\rm Re}\lambda_5+\nu
\end{pmatrix} \ ,
\end{equation}
where $\lambda_{345}=\lambda_3 + \lambda_4 + {\rm Re}\lambda_5$. We define an orthogonal rotation matrix $R$ to diagonalize the above mass matrix, with $R{\cal M}^2R^{\rm T}={\rm diag}(m_{h_1}^2,m_{h_2}^2,m_{h_3}^2)$. Generally, the matrix $R$ can be parametrized as~\cite{WahabElKaffas:2007xd, Hayashi:1994xf}
\begin{eqnarray}\label{R}
R = R_{23} (\alpha_c) R_{13} (\alpha_b) R_{12} (\alpha + \pi/2) =\begin{pmatrix}
-s_{\alpha}c_{\alpha_b} & c_{\alpha}c_{\alpha_b} & s_{\alpha_b} \\
s_{\alpha}s_{\alpha_b}s_{\alpha_c} - c_{\alpha}c_{\alpha_c} & -s_{\alpha}c_{\alpha_c} - c_{\alpha}s_{\alpha_b}s_{\alpha_c} & c_{\alpha_b}s_{\alpha_c} \\
s_{\alpha}s_{\alpha_b}c_{\alpha_c} + c_{\alpha}s_{\alpha_c} & s_{\alpha}s_{\alpha_c} - c_{\alpha}s_{\alpha_b}c_{\alpha_c} & c_{\alpha_b}c_{\alpha_c}
\end{pmatrix} \ .
\end{eqnarray}
Both $\alpha_b$ and $\alpha_c$ are CP violating mixing angles in the Higgs sector that depend implicitly on $\delta_1$. 
In this convention, the significance of the $\alpha$ and $\beta$ angles are the same as in the minimal supersymmetric Standard Model, and the interactions with quarks of lightest Higgs state, $h_1$, depends only on one CP violating angle $\alpha_b$. The mass and CP eigenstates are related via $(H_1^0, H_2^0, A^0) = (h_1, h_2, h_3) R$. As we discuss below, $\alpha_c$ is determined once $\alpha_b$, $\alpha$, $\beta$, and the neutral scalar masses are specified. We will, thus, utilize $\alpha_b$ rather than $\delta_1$ to characterize the effects of CPV in the potential. 

\subsection{Interactions}
For phenomenological analysis, we are interested in interactions of the scalar sector with the other SM particles. 
After EWSB, the couplings of the neutral scalars with fermions and gauge bosons can be parametrized generally as~\cite{Haber:1978jt}
\begin{eqnarray}\label{Lcpv}
\mathcal{L} = - \frac{m_f}{v} h_i  \left( c_{f,i} \bar f f+ \tilde c_{f,i} \bar f i\gamma_5 f  \right) + a_i h_i \left( \frac{2m_W^2}{v} W_\mu W^\mu + \frac{m_Z^2}{v} Z_\mu Z^\mu \right) \ ,
\end{eqnarray}
where $i=1,2,3$ and where we have allowed for only flavor diagonal couplings. To arrive at the couplings, we extend the $Z_2$ symmetry of the scalar potential by making the following assignments to the fermions:
\bea
Q_L\to Q_L\, \quad u_R\to u_R\, \quad d_R\to d_R, & \mathrm{Type\ I}\ \ \ ,\\
Q_L\to Q_L\, \quad u_R\to u_R\, \quad d_R\to -d_R, & \mathrm{Type\ II}\ \ \ .
\eea
One may make similar assignments for the leptons. 
The resulting Yukawa interactions before EWSB are
\begin{eqnarray}\label{cff}
\mathcal{L}_{\rm I} = - Y_U \overline Q_L (i\tau_2) \phi_2^* u_R - Y_D \overline Q_L \phi_2 d_R + {\rm h.c.} \ , \\
\mathcal{L}_{\rm II} = - Y_U \overline Q_L (i\tau_2) \phi_2^* u_R - Y_D \overline Q_L \phi_1 d_R + {\rm h.c.} \ .
\end{eqnarray}
Note that $\mathcal{L}_{I,II}$ satisfy the GW criterion for the absence of tree-level FCNCs. 

For each of the two types of models, we solve for $c_f$, $\tilde c_f$, $a$ in terms of $\beta$ and the orthogonal matrix $R$:
\begin{eqnarray}\label{Hcouplings}
\centering{\begin{tabular}{c|c|c|c|c|c}
\hline
&  $c_{t,i}$ & $c_{b,i}$ & $\tilde c_{t,i}$ & $\tilde c_{b,i}$ & $a_i$ \\
\hline
Type I & $R_{i2}/\sin\beta$ & $R_{i2}/\sin\beta$ & $-R_{i3}\cot\beta$ & $R_{i3}\cot\beta$ & 
$R_{i2}\sin\beta+R_{i1}\cos\beta$ \\
\hline
Type II & $R_{i2}/\sin\beta$ & $R_{i1}/\cos\beta$ & $-R_{i3}\cot\beta$ & $-R_{i3}\tan\beta$ & 
$R_{i2}\sin\beta+R_{i1}\cos\beta$ \\
\hline
\end{tabular}}
\end{eqnarray}
where all the up (down) type fermions have the universal rescaled couplings as the top (bottom) quark apart from the overall factor of the quark mass. 

The charged Higgs-fermion interactions are, respectively,
\begin{eqnarray}\label{flavor}
\mathcal{L}_{\bar f f' H^\pm} =\left\{\begin{array}{ll} V_{ij} \cot\beta \bar u_i [ m_{u_i}  (1-\gamma_5) + m_{d_j} (1+\gamma_5)] d_j H^+ + {\rm h.c.} & \hspace{1cm} {\rm type\ I} \\
V_{ij} \bar u_i [ m_{u_i} \cot\beta (1-\gamma_5) - m_{d_j} \tan\beta (1+\gamma_5)] d_j H^+ + {\rm h.c.} & \hspace{1cm} {\rm type\ II} 
\end{array} \right.
\end{eqnarray}
where $V$ stands for the CKM matrix for quark mixings.

The trilinear interactions between charged and neutral scalars, relevant for the scalar sector contribution to EDMs, are of the form
\begin{eqnarray}
\mathcal{L}_{H^\pm} = - \bar\lambda_i v h_i H^+ H^- \ ,
\end{eqnarray}
where $h_i$ and $H^\pm$ are mass eigenstates, and 
\begin{eqnarray}\label{lamtil}
\bar\lambda_i 
&=& R_{i1} \cdot \left( \lambda_3 \cos^2\beta + (\lambda_1 - \lambda_4 - {\rm Re}\lambda_5) \sin^2\beta \right) \cos\beta \nonumber \\ 
&+& R_{i2} \cdot \left( \lambda_3 \sin^2\beta + (\lambda_2 - \lambda_4 - {\rm Re}\lambda_5) \cos^2\beta \right) \sin \beta + R_{i3} \cdot {\rm Im} \lambda_5 \, \sin\beta \cos\beta \ .
\end{eqnarray}
We do not write down the corresponding quartic terms as they are not needed for our phenomenological analysis.

\subsection{Phenomenological parameters}

From the Higgs potential Eq.~(\ref{Eq:gko-pot}), it is possible to solve for the Higgs doublet VEVs as well as  the scalar masses and mixing angles, which are more directly related to observation. Since the aim of this work is to arrive at the phenomenological constraints on the parameter space of 2HDM, it is useful to translate these constraints into those on phenomenologically-relevant parameters. The latter set includes the masses, mixing angles, and the parameter $\nu$ introduced in Eq.~(\ref{MC}) and whose significance we discuss below. The following table summarizes two sets of parameters (all real).
\begin{eqnarray}\label{param}
\begin{tabular}{c|c}
\hline
Potential parameters & Phenomenological parameters \\
\hline
$\lambda_1$, $\lambda_2$, $\lambda_3$, $\lambda_4$, ${\rm Re}\lambda_5$, ${\rm Im}\lambda_5$  & $v$, {$\tan\beta$}, $\nu$, $\alpha$, $\alpha_b$, $\alpha_c$ \\
$m_{11}^2$, $m_{22}^2$, ${\rm Re}m_{12}^2$, ${\rm Im}m_{12}^2$  & $m_{h_1}$, $m_{h_2}$, $m_{h_3}$, $m_{H^+}$ \\
\hline
\end{tabular}
\end{eqnarray}
Each set has 10 parameters, and it would appear to be possible solve one set of parameters from the other. However, the minimization conditions for  the $A^0_k$ in Eq.~(\ref{mini3}) imply that there exists only one independent CPV phase and hence, that the CPV mixing angles $\alpha_b$ and $\alpha_c$ are not independent. As we show below, one may solve for $\alpha_c$ ($\alpha_b$) in terms of $\alpha_b$ ($\alpha_c$), the physical neutral scalar masses, $\alpha$ and $\beta$.

Two additional remarks are in order. First, the phenomenological significance of the parameter $\nu$ is that it controls the mass scale of the second Higgs doublet. In the decoupling limit wherein one reverts to the SM, one has $\nu\gg 1$. Eqs.~(\ref{MC}) and (\ref{MM}) then imply that $H^\pm$, $h_3\approx A^0$ and the linear combination $h_2\approx \sin\beta H_1^0 -\cos\beta H_2^0$ also decouple with an approximately common mass $\nu$. The resulting low energy theory contains only one CP even scalar $h_1$, which is the SM Higgs boson. In the same decoupling limit, we also have $\alpha_{b,c}\to0$, $\alpha\to\beta-\pi/2$. Away from the decoupling limit, both doublets are at the electroweak scale, and we have to treat $\nu$ as an independent input parameter. 

Second, it is useful to consider the CP conserving limit, with a real Higgs potential, i.e., ${\rm Im}\lambda_5=0$ and ${\rm Im}m_{12}^2=0$. In absence of SCPV, $\xi=0$, and the matrix~(\ref{MM}) is block diagonalized with vanishing ${\cal M}^2_{13}$ and ${\cal M}^2_{23}$ elements. In this regime, one has $\alpha_b=\alpha_c=0$, and
the independent parameters become
\begin{eqnarray}
\begin{tabular}{c|c}
\hline
Potential parameters & Phenomenological parameters (no CPV) \\
\hline
$\lambda_1$, $\lambda_2$, $\lambda_3$, $\lambda_4$, $\lambda_5$  & $v$, $\tan\beta$, $\alpha$ \\
$m_{11}^2$, $m_{22}^2$, $m_{12}^2$  & $m_{h_1}$, $m_{h_2}$, $m_{h_3}$, $m_{H^+}$ \\
\hline
\end{tabular}
\end{eqnarray}
Although there exist eight potential parameters in this case, the condition of no SCPV reduces the number of independent parameters to seven, which one may choose to be those in the right hand column of the table.

\bigskip
For the general scenario that allows for CPV, it is useful to write down  the relationships between the phenomenological parameters and those in the potential:
\begin{eqnarray}
\tan\beta &=& \frac{(m_{h_2}^2 -m_{h_3}^2) \cos\alpha_c \sin\alpha_c + (m_{h_1}^2 -m_{h_2}^2 \sin^2\alpha_c-m_{h_3}^2 \cos^2\alpha_c) \tan\alpha \sin\alpha_b}
{(m_{h_2}^2 -m_{h_3}^2) \tan\alpha \cos\alpha_c \sin\alpha_c - (m_{h_1}^2 -m_{h_2}^2 \sin^2\alpha_c-m_{h_3}^2 \cos^2\alpha_c) \sin\alpha_b} \ , \label{ab}\\
\label{eq:lambda1}
\lambda_1 &=& \frac{m_{h_1}^2 \sin^2\alpha \cos^2\alpha_b + m_{h_2}^2 R_{21}^2 
+ m_{h_3}^2 R_{31}^2}{v^2 \cos\beta^2} - \nu \tan^2\beta \ , \\
\lambda_2 &=& \frac{m_{h_1}^2 \cos^2\alpha \cos^2\alpha_b + m_{h_2}^2 R_{22}^2 
+ m_{h_3}^2 R_{32}^2}{v^2 \sin\beta^2} - \nu \cot^2\beta \ , \\
{\rm Re}\lambda_5 &=& \nu - \frac{m_{h_1}^2 \sin^2\alpha_b + \cos^2\alpha_b (m_{h_2}^2 \sin^2\alpha_c + m_{h_3}^2 \cos^2\alpha_c)}{v^2} \ , \\
\lambda_4 &=& 2 \nu - {\rm Re}\lambda_5 - \frac{2 m_{H^+}^2}{v^2} \ , \\
\lambda_3 &=& \nu - \frac{m_{h_1}^2 \sin\alpha \cos\alpha \cos^2\alpha_b - m_{h_2}^2R_{21}R_{22} - m_{h_3}^2R_{31}R_{32}}{v^2\sin\beta\cos\beta} - \lambda_4 - {\rm Re}\lambda_5 \ , \\
{\rm Im}\lambda_5 &=& \frac{2 \cos\alpha_b \left[ (m_{h_2}^2-m_{h_3}^2) \cos\alpha \sin\alpha_c \cos\alpha_c +  (m_{h_1}^2 - m_{h_2}^2 \sin^2\alpha_c-m_{h_3}^2\cos^2\alpha_c)^2 \sin\alpha \sin\alpha_b \right]}{v^2 \sin\beta} \label{eq:imlambda5}  \ .
\end{eqnarray}
Note that Eq.~(\ref{ab}) implies that $\alpha_b$, $\alpha_c$, $\alpha$, $\beta$ and the neutral scalar masses are not all independent, as advertised. 
The remaining equations (\ref{eq:lambda1}--\ref{eq:imlambda5}), together with the minimization conditions (\ref{mini1}--\ref{mini3}),
can be used to solve for the 9 independent phenomenological parameters in Eq.~(\ref{param}).

In order to make the presence of only one independent CPV phase apparent, we chose to eliminate one of the two CPV mixing angles ($\alpha_c$) in terms of the other parameters, including the other CPV mixing angle ($\alpha_b$) that vanishes in the CP-conserving 2HDM and the remaining parameters that survive in the absence of CPV.

\subsubsection{Parameter ranges}

\begin{figure}[t!]
\centerline{\includegraphics[width=0.7\columnwidth]{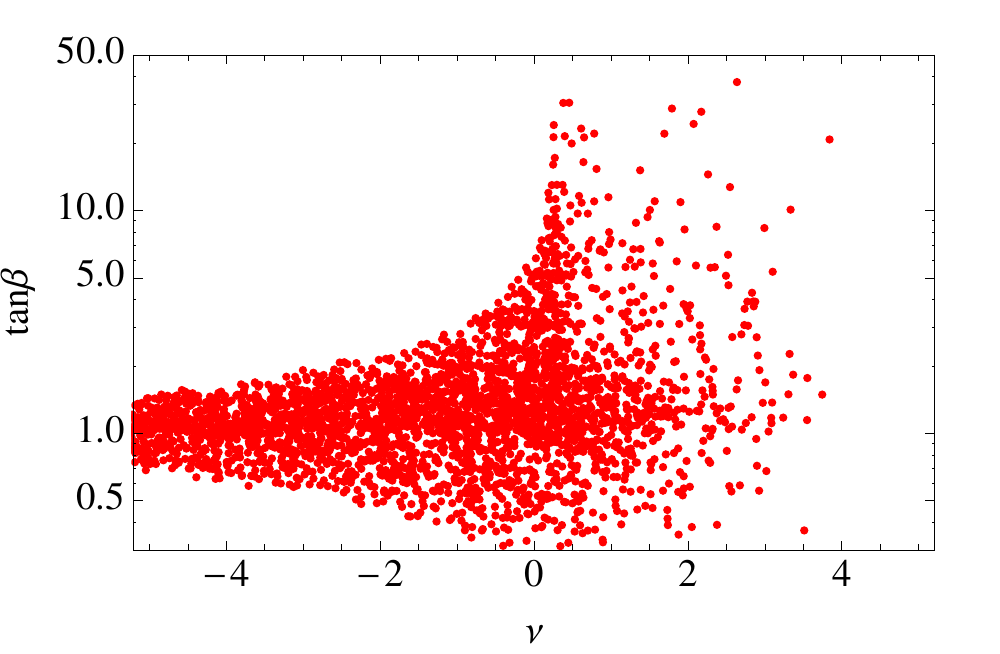}}
\caption{Theoretical constraints on the $\nu-\tan\beta$ parameter space.}\label{par}
\end{figure}

Theoretical constraints on the parameter space follow from requirements of stability of the electroweak vacuum and perturbativity~\cite{Grzadkowski:2013rza}. While the latter is not an absolute requirement for the validity of the theory, our phenomenological study relies on perturbative computations of observables, so we restrict our attention to the domain of na\"ive perturbativity, expressed in terms of the quartic couplings:
\begin{eqnarray}
0<\lambda_1 < 4\pi, \ \ \ 0<\lambda_2 < 4\pi, \ \ \ \lambda_3 > - \sqrt{\lambda_1 \lambda_2}, \ \ \ \lambda_3 + \lambda_4 - |\lambda_5| > - \sqrt{\lambda_1 \lambda_2} \ .
\label{eq:vacpert}
\end{eqnarray}
Using Eqs.~(\ref{eq:lambda1}-\ref{eq:imlambda5}) we translate these conditions into constraints on the phenomenological parameters. To illustrate, 
we take ${h_1}$ to be the 125\,GeV Higgs boson discovered at the LHC. For the ranges of other parameters we allow $m_{h_2}, m_{h_3} \in [125, 500]\,$GeV, $\alpha, \alpha_b \in [-\pi/2, \pi/2]$ (notice $\alpha_c$ is not independent as discussed above). The resulting region consistent with the conditions (\ref{eq:vacpert}) in the $\nu-\tan\beta$ plane is shown in Fig.~\ref{par}. 

\begin{figure}[t!]
\centerline{\includegraphics[width=0.7\columnwidth]{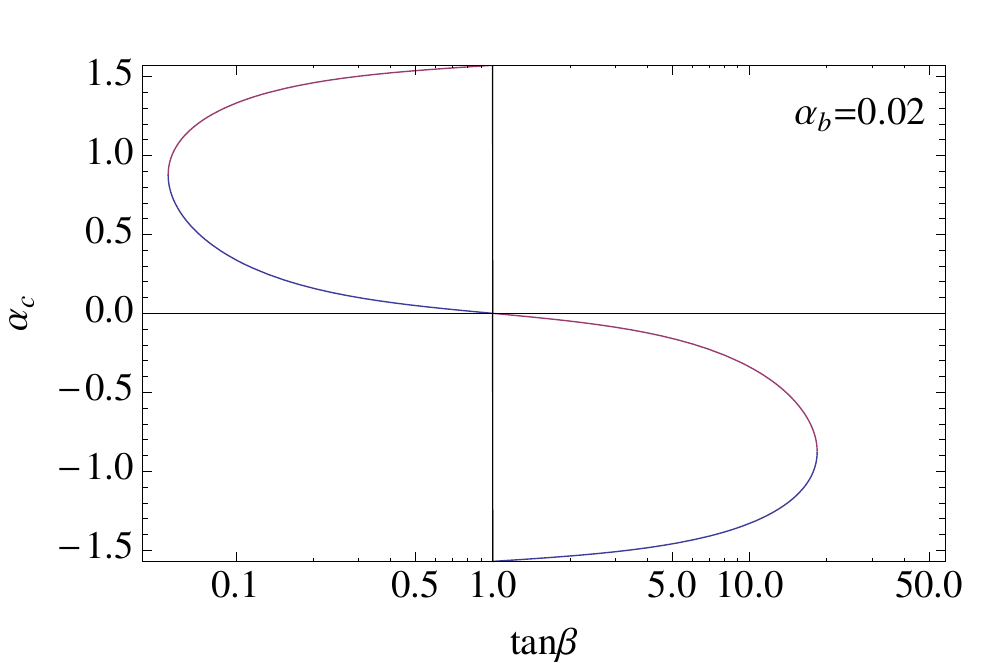}}
\caption{Solutions for $\alpha_c^+$ (blue), $\alpha_c^-$ (magenta) as a function of $\tan\beta$ using Eq.~(\ref{special2}) for fixed $\alpha_b=0.02$. The other parameters are $m_{H^+}=420\,$GeV, $m_{h_2}=\,400$\,GeV, $m_{h_3}=\,450$\,GeV and $\nu=1.0$.}\label{alp}
\end{figure}

As discussed in the fit to LHC Higgs data in Section \ref{sec:lhc} below, especially in the type-II 2HDM, the fit to the LHC data on the Higgs boson production and decay rates points to a strong correlation between the angles $\alpha$ and $\beta$, with $\alpha \approx \beta-\pi/2$ (see Figure \ref{alp}). If this is taken to be a constraint, we can further reduce the set of phenomenological input parameters to
\begin{eqnarray}
\alpha_b
, \ \ \tan\beta, \ \ m_{H^+}, \ \ m_{h_1}, \ \ m_{h_2}, \ \ m_{h_3}, \ \ \nu \ .
\end{eqnarray}
When either of the CPV mixing angles $\alpha_b$ or $\alpha_c$ is fixed, the other can be obtained from Eq.~(\ref{ab}):
\begin{eqnarray}\label{special1}
\alpha_b = \arcsin \left[\frac{(m_{h_2}^2-m_{h_3}^2) \sin2\alpha_c\tan2\beta}{2(m_{h_1}^2 - m_{h_2}^2 \sin^2\alpha_c-m_{h_3}^2 \cos^2\alpha_c)} \right] \ ,
\end{eqnarray}
or alternatively,
\begin{eqnarray}\label{special2}
\alpha_c^\pm = \arctan\left[ \frac{(m_2^2-m_3^2)\tan2\beta\pm \sqrt{ (m_2^2-m_3^2)^2\tan^22\beta - 4\sin^2\alpha_b (m_1^2-m_2^2)(m_1^2-m_3^2) }}{2(m_1^2-m_2^2)\sin\alpha_b} \right] \ .
\end{eqnarray}
For given $\alpha_b$, there are two solutions for $\alpha_c$. We find that they satisfy the relation
$\tan\alpha_c^+ \tan\alpha_c^- = ({m_{h_1}^2 - m_{h_3}^2})/({m_{h_1}^2-m_{h_2}^2})$,
which is approximately 1, in the limit $m_{h_1} \ll m_{h_2} \approx m_{h_3}$. Fig.~\ref{alp} illustrates the above relations between $\alpha_b$ and $\alpha_c$, for a set of sample parameters. One can see on the right panel that the expression for $\alpha_c^\pm$ contains a discontinuity at $\tan \beta=0$ (for fixed color, blue or magenta). For our phenomenological studies, we choose $\alpha_c^+$ for $\tan \beta < 1$ and $\alpha_c^-$ for $\tan \beta >1$, in order to avoid this discontinuity. Physically, our choice corresponds to $h_3$ being mostly CP-odd; $\alpha_c^-$ for $\tan \beta < 1$ and $\alpha_c^+$ for $\tan \beta > 1$ correspond to the case when $h_2$ is mostly CP-odd. We have observed that the choice between $\alpha_c^+$ and $\alpha_c^-$ does not make a qualitative difference in our conclusions discussed below.

\section{Observables}
\label{sec:obs}

\subsection{Event rates of all Higgs decay channels at LHC}
\label{sec:lhc}

In this work, we assume the light neutral scalar from the 2HDM is the 125 GeV Higgs boson discovered at the LHC. 
In the presence of CPV interactions in Eq.~(\ref{Lcpv}), the Higgs production and decay rates are modified as follows,
\begin{eqnarray} \label{rates}
\frac{\sigma_{gg\to h_1}}{\sigma^{\rm SM}_{gg\to h_1}} &=& \frac{\Gamma_{h_1 \to gg}}{\Gamma_{h \to gg}^{\rm SM}} \approx \frac{(1.03c_t - 0.06 c_b)^2 + (1.57\tilde c_t - 0.06 \tilde c_b)^2}{(1.03 - 0.06)^2} \ ,  \\
\frac{\Gamma_{h_1 \to \gamma\gamma}}{\Gamma_{h \to \gamma\gamma}^{\rm SM}} &\approx& \frac{(0.23 c_t - 1.04 a)^2+(0.35 \tilde c_t)^2}{(0.23 - 1.04)^2} \ , \\
\frac{\sigma_{VV\to h_1}}{\sigma^{\rm SM}_{VV\to h}} &=& \frac{\sigma_{V^*\to V h_1}}{\sigma^{\rm SM}_{V^*\to V h}} =\frac{\Gamma_{h_1 \to WW}}{\Gamma_{h \to WW}^{\rm SM}} = \frac{\Gamma_{h_1 \to ZZ}}{\Gamma_{h \to ZZ}^{\rm SM}} =a^2 \ , \\
\frac{\Gamma_{h_1 \to b\bar b}}{\Gamma_{h \to b \bar b}^{\rm SM}} &=& \frac{\Gamma_{h_1 \to \tau^+\tau^-}}{\Gamma_{h \to \tau^+\tau^-}^{\rm SM}} \approx  c_b^2 + \tilde c_b^2 \ .
\end{eqnarray}
The modified event rates will be constrained by the inclusive data in Higgs decay channels, which are summarized in Table.~\ref{data}. 
Since we are interested only in the couplings of $h_1$, all the above couplings $a$, $c_f$, $\tilde c_f$ can be expressed in terms of
only three parameters $\beta$, $\alpha$ and $\alpha_b$, where $\alpha_b$ is the CPV mixing angle.
In the presence of CP violation, the global fit to the combined results measured by the ATLAS~\cite{ATLAS} and CMS~\cite{CMS} collaborations have been performed in several previous works~\cite{Shu:2013uua, Freitas:2012kw, Celis:2013rcs,Djouadi:2013qya,Chang:2013cia}. In this work, we follow the same parametrization as~\cite{Shu:2013uua} and present the results for both type-I and II 2HDMs.

\begin{table}[h]
\begin{tabular}{c|c|c|c|c|c}
\hline
&  $\gamma\gamma$ & $WW$ & $ZZ$ & $Vbb$ & $\tau\tau$ \\
\hline
ATLAS & $1.6 \pm0.3$ & $1.0 \pm0.3$ & $1.5\pm0.4$ & $-0.4\pm1.0$ & $0.8\pm0.7$ \\
\hline
CMS &  $0.8\pm0.3$ & $0.8\pm0.2$ & $0.9\pm0.2$ & $1.3\pm0.6$ & $1.1\pm0.4$\\
\hline
\end{tabular}
\caption{LHC data on all measured Higgs decay channels from the ATLAS and CMS collaborations.}\label{data}
\end{table}

\subsection{T- and P-violating effective operators, RG running and matching} \label{running}

We now turn to the low-energy sector, focusing on the EDMs of the neutron, neutral atoms, and molecules\footnote{Technically speaking, the paramagnetic molecule response to the external field.}  that presently yield the most stringent constraints on flavor-diagonal CPV. The relevant time reversal- and parity-violating (TVPV) effective operators for our study are the elementary fermion EDMs, the quark chromo-EDMs (CEDM), and the Weinberg three gluon (or gluon CEDM) operators. The corresponding effective Lagrangian valid below the electroweak scale is
\begin{eqnarray}
\mathcal{L}_{\rm eff} & = &  -i\sum_f\, \frac{d_f}{2} \, \bar f \sigma_{\mu\nu} \gamma_5 f F^{\mu\nu} -i\sum_q\, \frac{{\tilde d}_q}{2} \bar q \sigma_{\mu\nu} \gamma_5 T^a q G^{a\mu\nu} +\frac{d_w}{6}  f^{abc} \epsilon^{\mu\nu\rho\sigma} G_{\mu\lambda}^a G_{\nu}^{b\, \lambda} G_{\rho\sigma}^c\\
 &\equiv & 
 \label{leff}
i\sum_f\, \frac{\delta_f}{\Lambda^2} m_{f} e \bar f \sigma_{\mu\nu} \gamma_5 f F^{\mu\nu} 
+ i \sum_q\, \frac{\tilde{\delta}_q}{\Lambda^2} m_{q} g_s \bar q \sigma_{\mu\nu} \gamma_5 T^a q G^{a\mu\nu}
+ \frac{C_{\tilde{G}}}{2 \Lambda^2} g_s f^{abc} \epsilon^{\mu\nu\rho\sigma} G_{\mu\lambda}^a G_{\nu}^{b\, \lambda} G_{\rho\sigma}^c \ ,
\end{eqnarray}
where we take the convention $\epsilon^{0123} = +1$, and in the first line we have expressed the effective operators in terms of dimensional coefficients, while in the second we have followed Ref.~\cite{Engel:2013lsa} and re-written them in
terms of the dimensionless quantities $\delta_f$, ${\tilde\delta}_q$, and $C_{\tilde G}$, the scale of physics beyond the Standard Model $\Lambda$, and the fermion masses. Doing so is consistent with the approach of a low-energy effective field theory, wherein one makes the relevant scales and their hierarchy explicit. On general grounds, one then expects the remaining dimensionless Wilson coefficients to be comparable in magnitude, all other considerations being equal. We note that in the versions of the 2HDM considered here, the dipole operators for a given fermion are naturally proportional to its Yukawa coupling, though in other BSM scenarios they need not necessarily be. Consequently, it is appropriate to scale out the fermion masses. The dimensionless EDM, $\delta_q$, quark CEDM, $\tilde{\delta}_q$, and gluon CEDM are  related to the usual definitions by~\cite{Engel:2013lsa}
\begin{eqnarray}
\delta_f \equiv -\frac{\Lambda^2 d_f^\gamma}{2 e Q_q m_q}, \ \ \ \ \ \tilde \delta_q \equiv -\frac{\Lambda^2 d_q^G}{2 m_q}, \ \ \ \ \ C_{\tilde G} = \frac{\Lambda^2
d_w}{3 g_s}\ \ \ \ .
\end{eqnarray}
Henceforth, we set $\Lambda = v$. The dominant contributions to these coefficients arise at two-loop level at the 2HDM scale $\Lambda\sim v$, and have been summarized in the Appendix~\ref{alledms}. 

\begin{figure}[h!]
\centerline{\includegraphics[width=0.35\columnwidth]{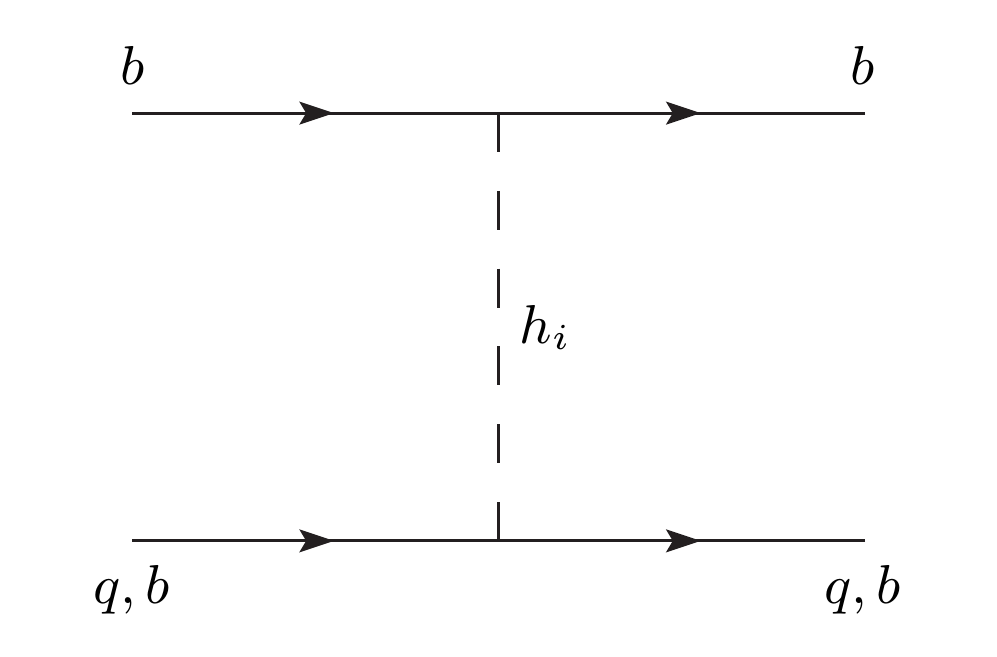}}
\caption{Tree level contribution to the P and T-odd four quark operators.}\label{tr}
\end{figure}

We also take account of the TVPV  four-quark operators that have an impact in the renormalization group (RG) running.
They arise in the 2HDM model at tree level from neutral Higgs exchange, as shown in Fig.~\ref{tr}.
We are particularly interested in the operators containing the bottom quark, whose coefficients are enhanced when $\tan\beta$ is large, thereby making a significant contribution in certain cases. Those involving only light quarks are suppressed by products of their small Yukawa couplings and are, therefore, neglected here. The operators under consideration are
\begin{eqnarray}\label{4qo}
\mathcal{L}^{4q}_{\rm eff} = \frac{C_4^b}{\Lambda^2} (\bar b b) (\bar b i\gamma_5 b) + \frac{\tilde C_1^{bq}}{\Lambda^2} (\bar b  b)(\bar q i\gamma_5 q) + \frac{\tilde C_1^{qb}}{\Lambda^2} (\bar q q) (\bar b i\gamma_5 b)
\ , 
\end{eqnarray}
where $q=u,d$. 
At the 2HDM scale $\Lambda = v$, these coefficients are
\begin{eqnarray}
C_4^{b} (\Lambda) = \sum_i\frac{m_b^2}{m_{h_i}^2} c_{b,i} \tilde c_{q,i}, \ \ \
\tilde C_1^{bq} (\Lambda) = \sum_i\frac{m_b m_q}{m_{h_i}^2} \tilde c_{b,i} c_{q,i}, \ \ \
\tilde C_1^{qb} (\Lambda) = \sum_i\frac{m_b m_q}{m_{h_i}^2} c_{b,i} \tilde c_{b,i} \ .
\end{eqnarray}

In order to calculate the neutron and atomic EDMs, we take account of the renormalization group running effect due to one-loop QCD corrections.
The Wilson coefficients in Eq.~(\ref{leff}) are evolved from $\Lambda$ down to the GeV scale, based on the RG equations (RGE)~\cite{Degrassi:2005zd,Hisano:2012cc,Dekens:2013zca} 
\begin{eqnarray}\label{RGE}
\frac{d}{d \ln \mu} \left(\frac{\delta_q}{Q_q}, \tilde{\delta}_q, - \frac{3 C_{\tilde{G}}}{2} \right) =  
\left(\frac{\delta_q}{Q_q}, \tilde{\delta}_q, - \frac{3 C_{\tilde{G}}}{2} \right) \cdot \frac{\alpha_S}{4 \pi}
\left(\begin{array}{ccc}
8 C_F & 0 & 0 \\
- 8 C_F & 16 C_F - 4N & 0 \\
0 & 2N & N + 2n_f + \beta_0
\end{array} \right) \ ,
\end{eqnarray}
where $q=u,d,b$, $N=3$, $C_F = (N^2-1)/(2N)=4/3$ and $\beta_0 = (11N-2n_f)/3$. 

Between the 2HDM scale $\Lambda$ and  $m_b$, we use the $n_f=5$ version of the above RGE. In addition, there are contributions through mixing from the four-quark operators in Eq.~(\ref{4qo}). In particular, the coefficient $C_4^b$ mixes with, and contributes to, the b-quark CEDM, and captures the leading logarithmic terms of the one-loop result~\cite{Barger:1996jc}. The coefficients $\tilde C_1^{bq}$, $\tilde C_1^{qb}$ also contribute to the light quark CEDM through RGE operator mixing as discussed in detail in~\cite{Hisano:2012cc}. This reproduces the leading logarithmic terms in the Barr-Zee type contribution to the CEDM with a b-quark in the upper loop.
In this calculation, we keep only the leading logarithmic terms that make additional contributions to  the CEDMs of bottom and light quarks at the matching scale $\mu=m_b$:
\begin{eqnarray}
\Delta \tilde{\delta}_b (m_b) &\approx& \frac{1}{8\pi^2} C_4^{b} (\Lambda) \log\frac{\Lambda}{m_b} \ , \label{shiftbcedm} \\
\Delta \tilde{\delta}_q(m_b) &\approx&  \frac{g_s^2}{64\pi^4} \frac{m_b}{m_q} (\tilde C_1^{bq} + \tilde C_1^{qb}) \left(\log\frac{\Lambda}{m_b}\right)^2  \ .
\end{eqnarray}
At the same scale, the bottom quark is integrated out, and its CEDM makes a shift to the Weinberg operator~\cite{Boyd:1990bx, Hisano:2012cc},
\begin{eqnarray}
\Delta C_{\tilde{G}} (m_b) = \frac{\alpha_S(m_b)}{12\pi} \tilde{\delta}_b(m_b) \ ,
\end{eqnarray}
where the b-quark CEDM $\tilde{\delta}_b(m_b)$ at the $m_b$ scale includes both the top quark contribution Eq.~(\ref{A4}) which evolves under the RGE, and the shift (\ref{shiftbcedm}).

After taking these renormalization effects,
the coefficients $\delta_q$, $\tilde{\delta}_q^G$ and $C_{\tilde{G}}$ are further evolved down to the GeV scale according to RGE with 4 or 3 flavors, for the interval above or below the charm quark mass scale, respectively.

\subsection{Current and future EDM constraints} 
\label{sec:edmconstraints}

We now analyze the constraints implied by EDM search null results considering, in turn, paramagnetic atoms and molecules, the neutron, and diamagnetic atoms. Within the context of the flavor conserving 2HDMs, constraints from the paramagnetic systems translate into limits on the electron EDM ($d_e$), while in a model-independent analysis paramagnetic results bound a linear combination of  $d_e$ and a dimension-six semileptonic interaction. The neutron EDM ($d_n$) is sensitive primarily to the quark EDM and CEDMs as well as the CPV three gluon operator, while the CPV four-(light)quark operator contributions are subdominant in the 2HDM context. For the diamagnetic systems, such as $^{199}$Hg, the quark EDM contribution is in general relatively suppressed, as is a dimension six semileptonic tensor interaction that can be more important in other contexts apart from the 2HDM. The quark CEDM and three gluon operators, thus, are the most significant for the diamagnetic systems in the 2HDM. 

\vskip 0.15in

\noindent{\bf Electron EDM.} \ \ Currently, the electron EDM is most strongly constrained by the ACME experiment~\cite{Baron:2013eja}, which searched for an energy shift of ThO molecules due to an external electric field. The external field induces the spin of the unpaired electron to lie along the intermolecular axis, sampling the large internal electric field associated with the polar molecule. The measured energy shift is also sensitive to the TVPV electron-nucleon interaction
\begin{eqnarray}
\label{eq:semilept}
\mathcal{L}_{eN}^{\rm eff} = -\frac{G_F}{\sqrt2} C_S^{(0)} \bar e i\gamma_5 e \bar N N +\cdots \ \ \ ,
\end{eqnarray}
where the  \lq\lq$+\cdots$" in Eq.~(\ref{eq:semilept}) denote subleading semileptonic interactions and where the leading term
arises from the four fermion operators \cite{Engel:2013lsa}
\begin{equation}
\left[{\rm Im} C_{ledq} (\bar e i\gamma_5 e) (\bar d d) -  {\rm Im} C^{(1)}_{lequ} (\bar e i\gamma_5 e) (\bar u u) \right]/(2v^2)\ \ \ .
\end{equation}
 In the 2HDMs considered in this work, these four-fermion operators are obtained by integrating out the neutral Higgs bosons at tree-level,
\begin{eqnarray}
C_S^{(0)} = - g_s^{(0)} \left( {\rm Im} C_{ledq} - {\rm Im} C^{(1)}_{lequ} \right) =  - 2 g_s^{(0)} \sum_{i=1}^3 \frac{m_e}{m_{h_i}^2} \left( m_d \tilde c_{e,i} c_{d,i} + m_u \tilde c_{e,i} c_{u,i} \right) \ ,
\end{eqnarray}
where $g_s^{(0)}$ is the isoscalar nucleon scalar density form factor at zero momentum transfer (also known as the \lq\lq $\sigma$-term").
The ACME experiment gives the constraint~\cite{Baron:2013eja}
\begin{eqnarray}\label{acmelimit}
\left| \mathcal{E}_{\rm eff} d_e + W_S C_S^{(0)} \right| < 7.0\times10^{-18}\,{\rm eV} \ ,
\end{eqnarray}
where the effective field experienced by the unpaired electron is $\mathcal{E}_{\rm eff}=84\,{\rm GV/cm}$ and where $W_S=1.2\times10^{-9}\,{\rm eV}$.
Since $C_S^{(0)}$ is proportional to the product of the electron mass $m_e$ and light quark masses $m_{u,d}$ one may safely neglect the semileptonic interaction and translate Eq.~(\ref{acmelimit}) into a bound on $d_e$, or equivalently, $\delta_e$.
 
\bigskip
\noindent{\bf Neutron EDM.} \ \ The dependence of the neutron EDM on the leading nonleptoic CPV operators in the 2HDM 
is given by~\cite{Engel:2013lsa} 
\begin{eqnarray}
d_n = \left( e \zeta_n^u \delta_u + e \zeta_n^d \delta_d \right) + 
\left( e \tilde \zeta_n^u \tilde\delta_u + e \tilde\zeta_n^d \tilde\delta_d \right) + \beta_n^{G} C_{\tilde{G}} \ ,
\end{eqnarray}
where we have set $\Lambda=v$ as indicated earlier and 
where the central values~\cite{Engel:2013lsa} for the hadronic matrix elements are $\zeta_n^u=0.82\times10^{-8}$, $\zeta_n^d=-3.3\times10^{-8}$, $\tilde\zeta_n^u=0.82\times10^{-8}$, $\tilde \zeta_n^d=1.63\times10^{-8}$ and $\beta_n^{G}=2 \times 10^{-20} \,e\,{\rm cm}$.
The experimental upper bound on neutron EDM is~\cite{Baker:2006ts}
\begin{eqnarray}
d_n < 2.9\times 10^{-26} \,e\,{\rm cm} \ .
\end{eqnarray}

\bigskip
\noindent{\bf Diamagnetic Atom EDMs.} \ \ At present, the most stringent EDM limit has been obtained on the ${\rm ^{199}Hg}$ atom (see below). Efforts are underway to increase the sensitivity of this EDM search while other groups are pursuing searches for the EDMs of other diamagnetic atoms, including ${\rm ^{225}Ra}$ and $^{129}$Xe (for a discussion, see {\em e.g.} Ref.~\cite{Kumar:2013qya}). In what follows, we will consider the present ${\rm ^{199}Hg}$
constraint as well as the prospective impact of future ${\rm ^{199}Hg}$ and ${\rm ^{225}Ra}$ searches.
Diamagnetic atom EDMs arise primarily from their nuclear Schiff moments and a tensor semileptonic interaction. In the 2HDMs, the latter is suppressed by the same light fermion Yukawa factors that suppress $C_S^{(0)}$. The Schiff moment is generated by long-range, pion-exchange mediated P- and T-violating nucleon-nucleon interactions, where one vertex involves the P- and T-conserving strong $\pi NN$ coupling and the second consists of TVPV interaction:
\begin{equation}
\label{eq:piNN1}
\mathcal{L}_{\pi NN}^\mathrm{TVPV} = {\bar N}\left[ \bar g_{\pi}^{(0)} {\vec\tau}\cdot{\vec \pi} +  \bar g_{\pi}^{(1)} \pi^0 +\bar g_{\pi}^{(2)}(2\tau_3\pi^0-{\vec\tau}\cdot{\vec \pi})\right]N\ ,
\end{equation}
where the terms on the right hand side correspond to isoscalar, isovector, and isotensor channels, respectively. In general, the isotensor coupling $\bar g_{\pi}^{(2)}$ is suppressed with respect to the other two~\cite{Engel:2013lsa}, so we include only the latter in our analysis. Denoting the nuclear Schiff moment as $S$, one has~\cite{Engel:2013lsa},
\begin{eqnarray}
d_{\rm Hg} = \kappa_S S \approx \kappa_S \frac{2 m_N g_A}{F_\pi} \left(a_0 \bar g_{\pi}^{(0)} + a_1 \bar g_{\pi}^{(1)}\right) \ ,
\end{eqnarray}
where $g_A\approx 1.26$ and $F_\pi=186\,$MeV.  For $^{199}$Hg, we take the central values for the nuclear matrix elements from Ref.~\cite{Engel:2013lsa}: $a_0=0.01\,e\,{\rm fm}^3$, $a_1=\pm0.02\,e\,{\rm fm}^3$, while the atomic sensitivity coefficient is $\kappa_S=-2.8\times10^{-4}\,{\rm fm}^{-2}$ \cite{oai:arXiv.org:hep-ph/0203202}\footnote{We note that Eq. (5.181) of Ref.~\cite{Engel:2013lsa} omitted the minus sign on this value of $\kappa_S$.}. 
For Radium, the central values are $a_0=-1.5\,e\,{\rm fm}^3$, $a_1=6.0\,e\,{\rm fm}^3$, and $\kappa_S=-8.5\times10^{-4}\,{\rm fm}^{-2}$~\cite{oai:arXiv.org:hep-ph/0203202}.

At the hadronic level, the TVPV coefficients $\bar g_{\pi}^{(0,1)}$ arise from quark CEDMs and the Weinberg operators~\cite{Engel:2013lsa}, 
\begin{eqnarray}
\bar g_{\pi}^{(0)} = \tilde \eta_{(0)} ({\tilde\delta}_u^G + {\tilde\delta}_d^G) + \gamma^{\tilde{G}}_{(0)} C_{\tilde{G}}, \ \ \ \bar g_{\pi}^{(1)} = \tilde \eta_{(1)} ({\tilde\delta}_u^G - {\tilde\delta}_d^G) + \gamma^{\tilde{G}}_{(1)} C_{\tilde{G}} \ ,
\end{eqnarray}
where the hadronic matrix elements are $\tilde \eta_{(0)}=-2\times10^{-7}$, $\tilde \eta_{(1)}=-4\times10^{-7}$, $\gamma^{\tilde{G}}_{(0)}\approx \gamma^{\tilde{G}}_{(1)} = 2\times 10^{-6}$. There is also a contribution to $\bar g_{\pi}^{(i)}$ from the PV four quark operators, which we find to be unimportant for the 2HDMs here.

The current experimental upper bound on Mercury EDM is~\cite{Griffith:2009zz}
\begin{eqnarray}
d_{\rm Hg} < 3.1\times 10^{-29} \,e\,{\rm cm} \ ,
\end{eqnarray}
and we will use a conservative future sensitivity for the Radium EDM~\cite{Kumar:2013qya}
\begin{eqnarray}
d_{\rm Ra} < 10^{-27} \,e\,{\rm cm} \ .
\end{eqnarray}

\begin{figure}[t!]
\centerline{\includegraphics[width=1.0\columnwidth]{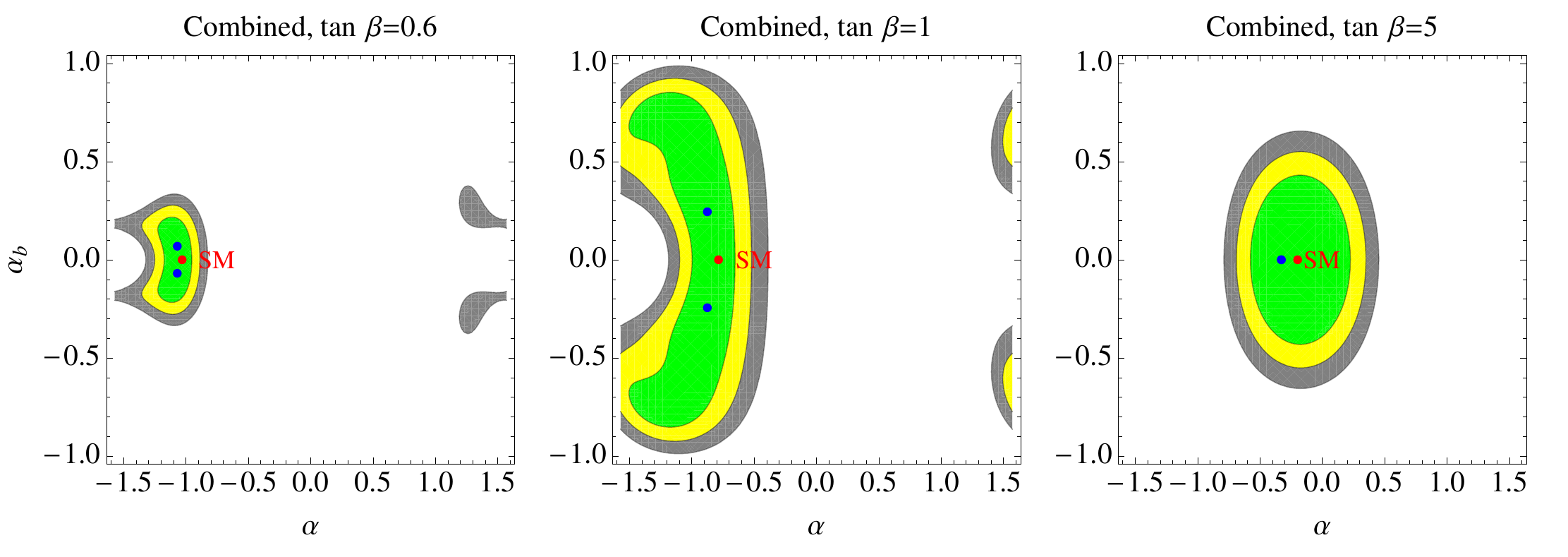}}
\centerline{\includegraphics[width=1.0\columnwidth]{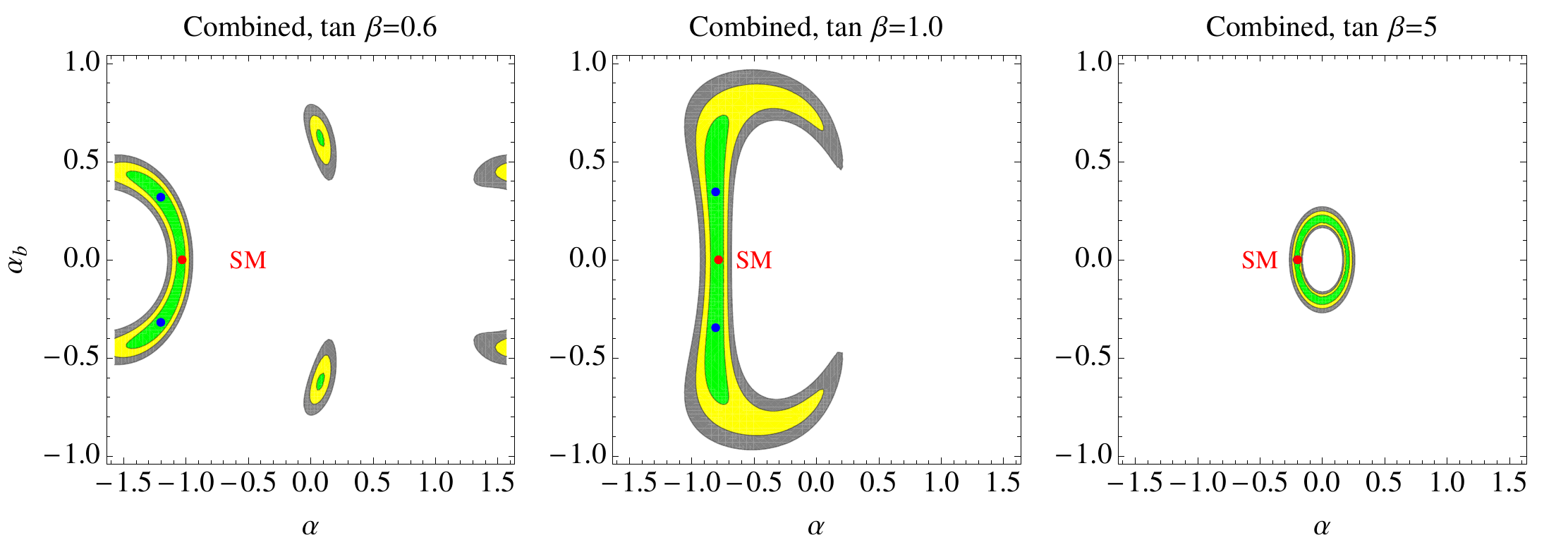}}
\caption{Global fit to the LHC Higgs data on the event rates given in Table~\ref{data}, for different values of $\tan\beta$. {\bf First row}: type-I model; {\bf Second row}: type-II model.
The other parameters are chosen to be $\alpha_c=0.05$, $m_{H^+}=420\,$GeV, $m_{h_2}=\,400$\,GeV and $m_{h_3}=\,450$\,GeV. 
The $1,2,3\sigma$ regions are in green, yellow, gray, and the best fit points are in blue. }\label{Fig:HiggsFit}
\end{figure}

\section{Results}
\label{sec:results}

\subsection{LHC Higgs Data Constraint}

We perform a global fit to the Higgs data given in Table~\ref{data}, where all the measured event rates are normalized to the SM predictions, and are of the form $\mu_f \pm \sigma_f$. The employed $\chi^2$ is defined as 
\begin{equation}
\chi^2 = \sum_f \frac{\left[{\sigma_{h_1} {\rm Br}_{h_1 \to f}}/({\sigma_{h_1}^{\rm SM} {\rm Br}_{h_1 \to f}^{\rm SM}}) - \mu_f\right]^2}{\sigma_f^2} \ ,
\end{equation}
where the sum goes through all the Higgs decay channels. The results are shown in Fig.~\ref{Fig:HiggsFit} in the two dimensional $\alpha_b-\alpha$ plane, for fixed values of $\tan\beta=0.6, 1, 5$, respectively. The $1,2,3\,\sigma$ regions are in green, yellow, gray, and the blue points represent where the best fit is found.

Clearly, the SM limit always gives a good fit, well within $1\sigma$. However, we find for low values of $\tan\beta \lesssim 1$, non-zero CP violation is preferred over the SM. This is mainly driven by the excess in the diphoton channel which still persists at ATLAS. We also find when $\tan\beta\sim1$ and $\alpha \approx \beta-\pi/2$, the largest possible $\alpha_b$ is allowed. In this case, the Higgs couplings defined in Eq.~(15) are $|c_t| \approx |c_b| \approx |a| \approx \cos\alpha_b$, and $|\tilde c_t|\approx |\tilde c_b|\approx \sin\alpha_b$. The resulting $\chi^2$ is a function of $\cos^2\alpha_b$, which when $\alpha_b\to0$, approaches $\chi^2_{\rm SM}$ for the SM case. Because the cosine function is rather ``flat" around the origin, there is substantial room for $\alpha_b$ to deviate from zero, while $\chi^2-\chi^2_{\rm SM}$ still remains small. 
The physical effect= of non-zero CP violation is to enhance the rate for $h_1\to\gamma\gamma$ and suppress the rate for $Vbb$ channel. At the same time, the Higgs total width is also slightly reduced.
At large $\tan\beta$, We find the effects of CPV are being enhanced, thus the good fit region shrinks with the increasing $\tan\beta$.
These facts suggest the best place to look for sizable possible CP violation angle is when $\tan\beta\approx1$. 

As indicated by Fig.~\ref{Fig:HiggsFit},  the LHC Higgs data imply strong constraints on the CP conserving angle $\alpha$, while the constraints on the CP violating angle $\alpha_b$ are generally relatively weaker. As we discuss below, EDM searches are more sensitive to the non-zero $\alpha_b$, but less sensitive to $\alpha$. In short, the Higgs studies and EDM searches provide complementary probes of the type-I and II 2HDMs.

\subsection{Electron EDM Constraint}

\begin{figure}[t!]
\centerline{\includegraphics[width=0.5\columnwidth]{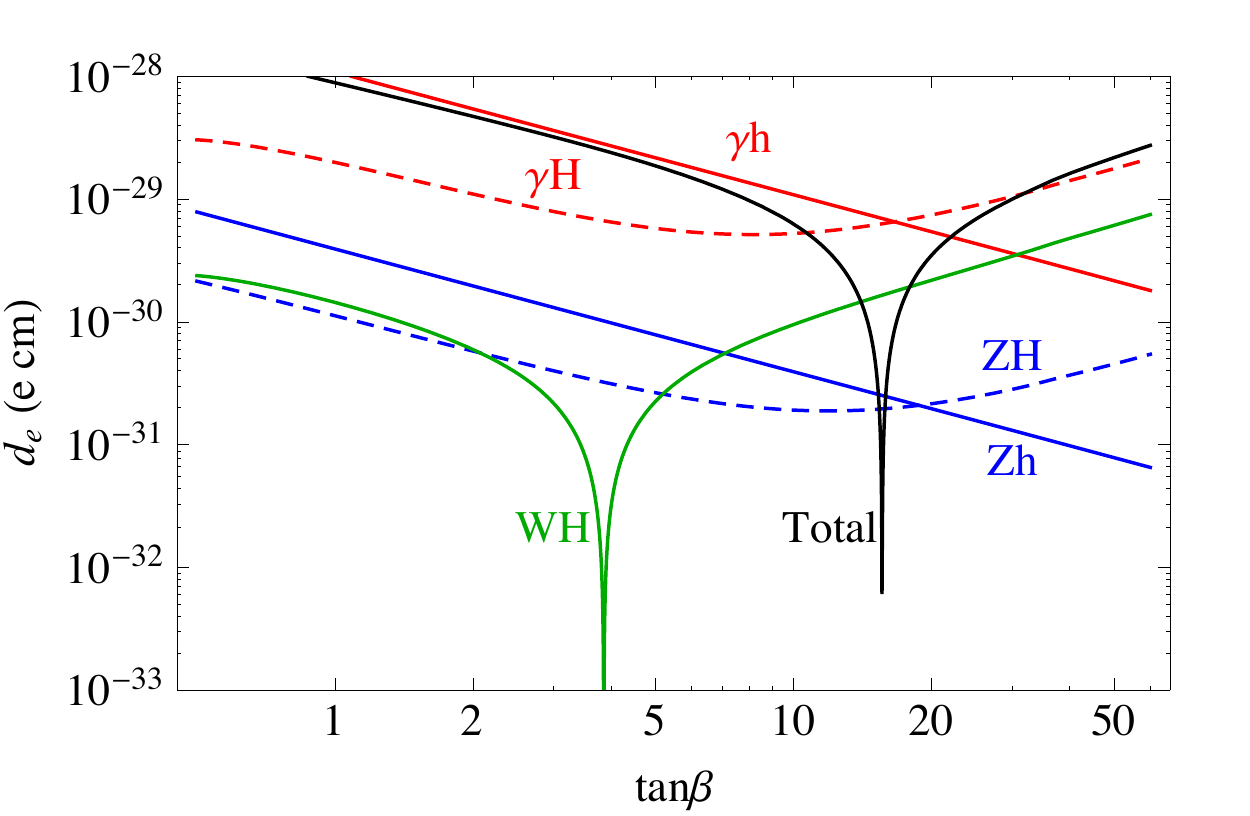}\includegraphics[width=0.5\columnwidth]{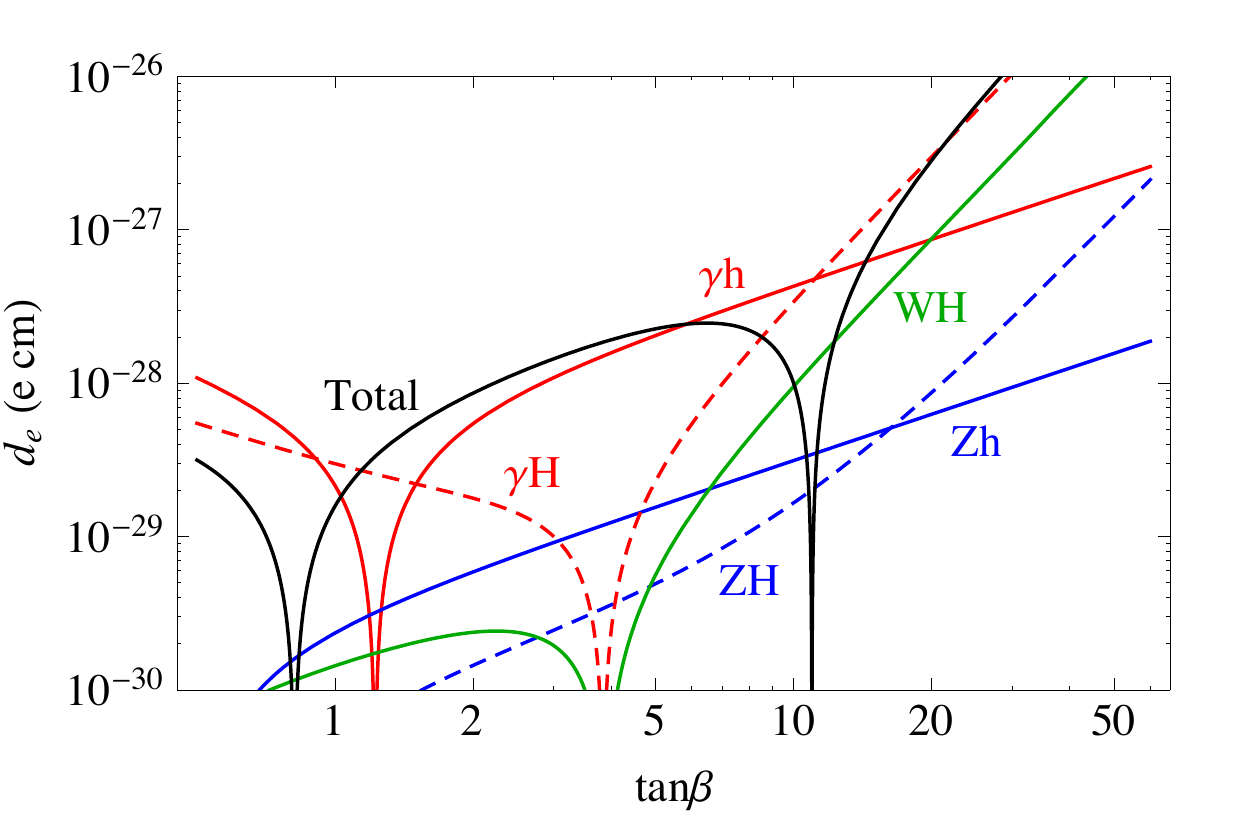}}
\caption{The anatomy of various contributions to the electron EDM in flavor conserving 2HDMs .{\bf Left}: type-I model; {\bf Right}: type-II model.
We plot the absolute values, so the dip in the curves implies a sign change.
Parameters are chosen to be $\alpha=\beta-\pi/2$, $\alpha_b=0.7\times10^{-2}$, $m_{H^+}=420\,$GeV, $m_{h_2}=\,400$\,GeV, $m_{h_3}=\,450$\,GeV and $\nu=1.0$. The parameter $\alpha_c$ is not independent and is obtained using Eq.~(\ref{special2}), and we note the two solutions $\alpha_c^\pm$ give very similar results here.  }\label{eEDManatomy}
\end{figure}

\begin{figure}[t!]
\centerline{\includegraphics[width=0.33\columnwidth]{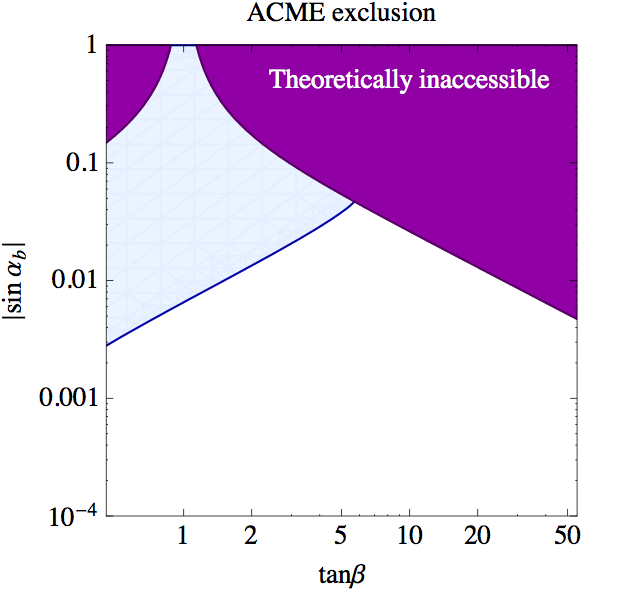}
\includegraphics[width=0.33\columnwidth]{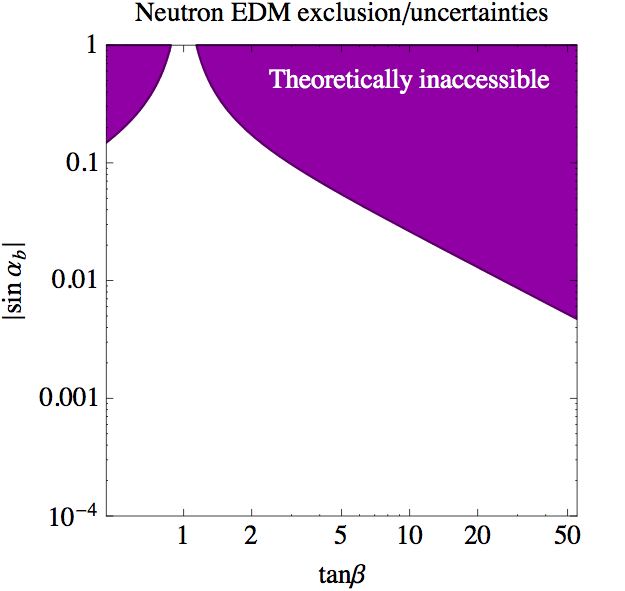}
\includegraphics[width=0.33\columnwidth]{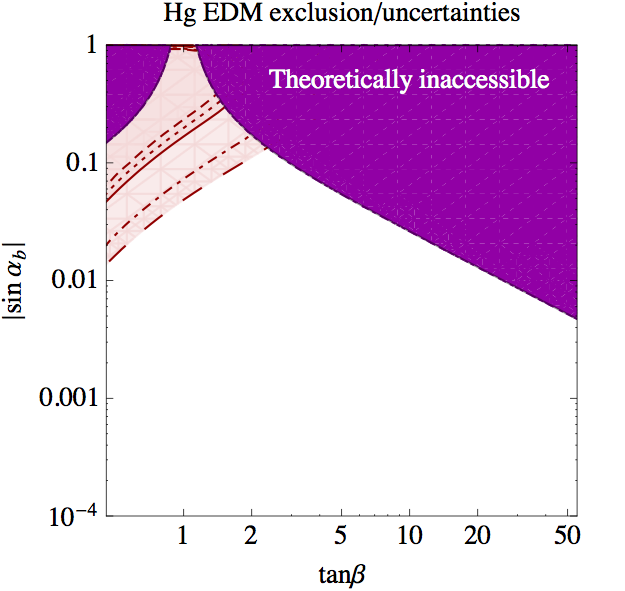}}
\centerline{\includegraphics[width=0.33\columnwidth]{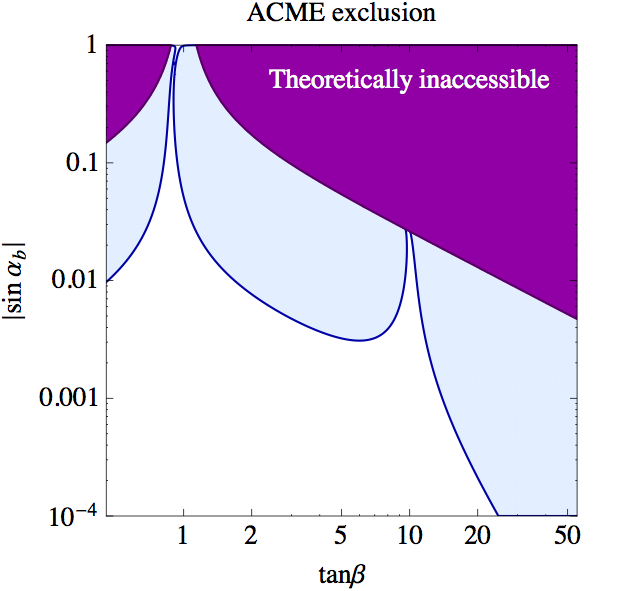}
\includegraphics[width=0.33\columnwidth]{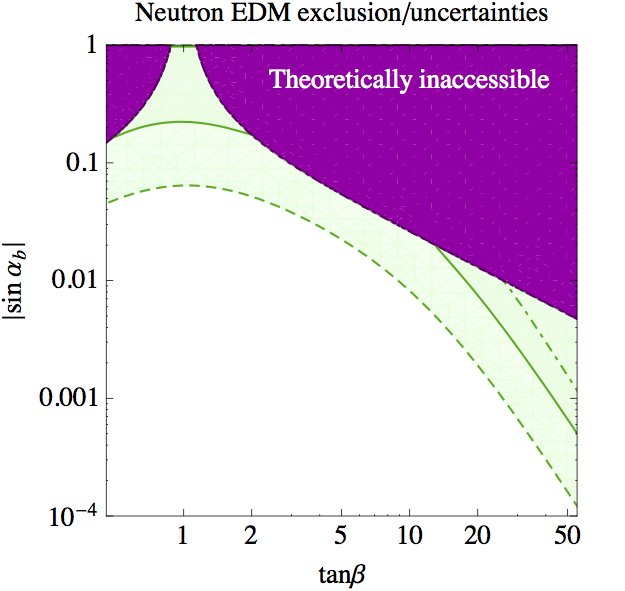}
\includegraphics[width=0.33\columnwidth]{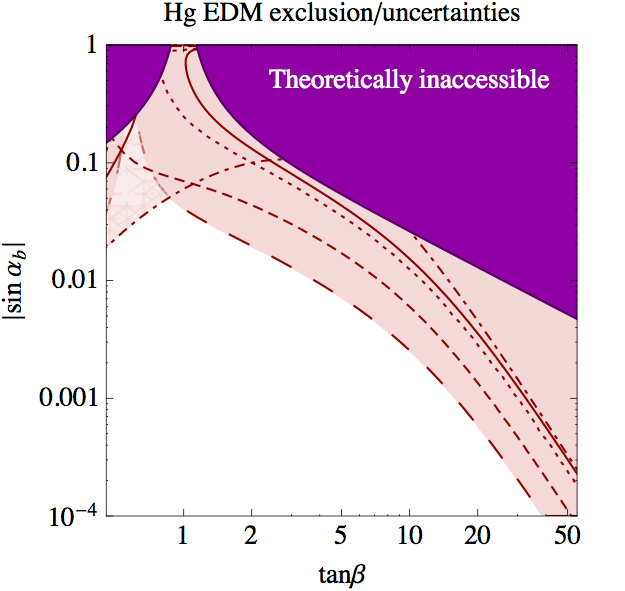}}
\caption{Current constraints from the electron EDM (left), neutron EDM (middle) and $^{199}$Hg EDM (right).{\bf First row}: type-I model; {\bf Second row}: type-II model. 
In all the plots, we have imposed the condition that $\alpha=\beta-\pi/2$. 
The other parameters are chosen to be $m_{H^+}=420\,$GeV, $m_{h_2}=\,400$\,GeV, $m_{h_3}=\,450$\,GeV and $\nu=1.0$.  
Again, $\alpha_c$ is a dependent parameter solved using Eq.~(\ref{special2}).
The purple region is theoretically not accessible because Eq.~(\ref{special2}) does not have a real solution.
For the neutron and Mercury EDMs, theoretical uncertainties from hadronic and nuclear matrix elements are reflected by different curves. 
For the neutron EDM, we vary one of the most important hadronic matrix elements: $\tilde \zeta_n^d = 1.63\times10^{-8}$ (solid, central value), $0.4\times10^{-8}$ (dot-dashed) and $4.0\times10^{-8}$ (dashed).  For the Mercury EDM, we take different sets of nuclear matrix element values: $a_0=0.01, a_1=0.02$ (solid, central value). $a_0=0.01, a_1=0.09$ (long-dashed), $a_0=0.01, a_1=-0.03$ (dashed), $a_0=0.005, a_1=0.02$ (dotted) and $a_0=0.05, a_1=0.02$ (dot-dashed).
} \label{bound&uncertainty}
\end{figure}

Drawing on the Wilson coefficients computed in Appendix \ref{alledms}; the hadronic, nuclear and atomic computations summarized in Section \ref{sec:edmconstraints}; and the present and prospective EDM search sensitivities, we present numerical results for the EDM constraints on the CPV parameter space. We give resulting constraints in the $\tan\beta$ vs. $\sin\alpha_b$ (Fig. \ref{bound&uncertainty}) as well as a breakdown, or anatomy, of the various contributions and their RG evolution as a function of $\tan\beta$ (Figs. \ref{eEDManatomy}, \ref{nEDManatomy}, \ref{pinnAnatomy}, and \ref{Fig:run}). We also discuss the implications of the constraints for the 2HDMs and the impact of the various hadronic and nuclear uncertainties.

\bigskip

We first consider the electron EDM, whose current limit is set by the ACME collaboration, $d_e < 10.25 \times 10^{-29} e$\,cm at 95\% confidence level.
Strictly speaking, ACME sets a bound on the linear combination of $d_e$ and a CP odd four fermion operator [see Eq.(\ref{acmelimit})]. In practice, we note the four fermion contribution is always subdominant in the flavor conserving 2HDMs.

In Fig.~\ref{eEDManatomy}, we plot the anatomy of various contributions [see Eq.~(\ref{A15}) in the appendix] to the electron EDM as functions of $\tan\beta$. 
Here we have fixed the CPV angle $\alpha_b=0.7\times10^{-2}$, then $\alpha_c$ is determined using Eq.~(\ref{special2}). 
We note the two solutions $\alpha_c^\pm$ give very similar plots. 
We fix the other parameters to be $m_{H^+}=420\,$GeV, $m_{h_2}=400\,$GeV, $m_{h_3}=450\,$GeV, $\nu=1.0$ and $\alpha=\beta-\pi/2$.\footnote{We have checked that our choice of masses is consistent with constraints from electroweak oblique parameters. General expressions for oblique parameters in 2HDM were given by Grimus et al.~\cite{Grimus:2008nb}.}\footnote{The charged Higgs mass is chosen to be $> 380\,$GeV in order to satisfy $B \rightarrow X_{s} \gamma$ bounds~\cite{Hermann:2012fc}.}\footnote{We note that the reported 3.4$\sigma$ deviation of the rate for ${\bar B}\to D^{(\ast)}\tau^-{\bar\nu}_\tau$~\cite{Lees:2012xj} from the SM prediction cannot be accommodated by the 2HDM scenario analyzed here. For the parameter space we consider, the discrepancy does not increase, though introduction of additional interactions would be needed to account for the difference. A study of the possibilities goes beyond the scope of the present study.}
We find the dominant contributions are always from $h\gamma\gamma$ or $H\gamma\gamma$ $(H=h_2, h_3)$ diagrams at low or high $\tan\beta$ regimes, respectively. Around $\tan\beta\approx10-20$, they have comparable magnitudes and opposite signs, leading to a cancellation in the total $d_e$. This cancellation leads to a sign change in $d_e$ but since we plot  $|d_e|$ it appears as a spike going toward zero.
The $H^\pm W^\mp \gamma$ contribution can give as large as 20\% corrections at large $\tan\beta$. The $hZ\gamma$ or $HZ\gamma$ contribution are always subdominant, because they are accidentally suppressed by the small $Zee$ vector coupling, proportional to $(1-4\sin^2\theta_W)$. 
 
The first column of Fig.~\ref{bound&uncertainty} shows the ACME experimental constraint, where the blue region is excluded. 
Here, in order to compare with the fit to LHC Higgs data results in Fig.~\ref{Fig:HiggsFit}, we have made the plots in the $\sin\alpha_b$--$\tan\beta$ plane.
Again, for given $\alpha_b=0.7\times10^{-2}$ there are two solutions for $\alpha_c$ from Eq.~(\ref{special2}): $\alpha_c^\pm$. We find that all the EDM constraints for both choices give very similar results. 
For the type II 2HDM, there are two cancellation regions in $\tan\beta$. The one near $\tan\beta\sim1$ was recently noticed in Ref.~\cite{Shu:2013uua, Ipek:2013iba}, which is due to the cancellation between the top quark and W-loop in the $h\gamma\gamma$ type diagrams.
The second region is near $\tan\beta\approx10-20$, is due to the cancellation between $h\gamma\gamma$ and $H\gamma\gamma$ contributions.
As we will show below, these cancellation regions can be closed when the neutron and mercury EDM limits are taken into account.
A generic feature is that for growing $\tan\beta$, the EDM constraints become weaker in the type-I 2HDM, but become stronger in the type-II 2HDM, which can be understood from the $\tan\beta$ dependences in Eq.~(\ref{Hcouplings}).

\subsection{Ineffectiveness of a Light-Higgs-Only Theory}

From the discussion of electron EDM, we have learned that the heavy Higgs contributions via $H\gamma\gamma$ and $H^\pm W^\mp\gamma$ diagrams make non-negligible contributions to the total EDM. They can even be dominant at large $\tan\beta\gtrsim20$. This example illustrates the ineffectiveness of the ``light Higgs effective theory", often performed as model independent analyses, which include the CPV effects only from the lightest Higgs (mass 125 GeV). The key point is that a CP violating Higgs sector usually contains more than one scalar at the electroweak scale, and all of them have CPV interactions in general. The total contribution therefore includes CPV effects from not only CP even-odd neutral scalar mixings, but also the CPV neutral-charged scalar interactions from the Higgs potential. This is necessarily model dependent. In this work, we have included the complete contributions to EDMs in the flavor-conserving (type-I and type-II) 2HDMs .

\subsection{Neutron EDM Constraint}

\begin{figure}[t!]
\centerline{\includegraphics[width=0.5\columnwidth]{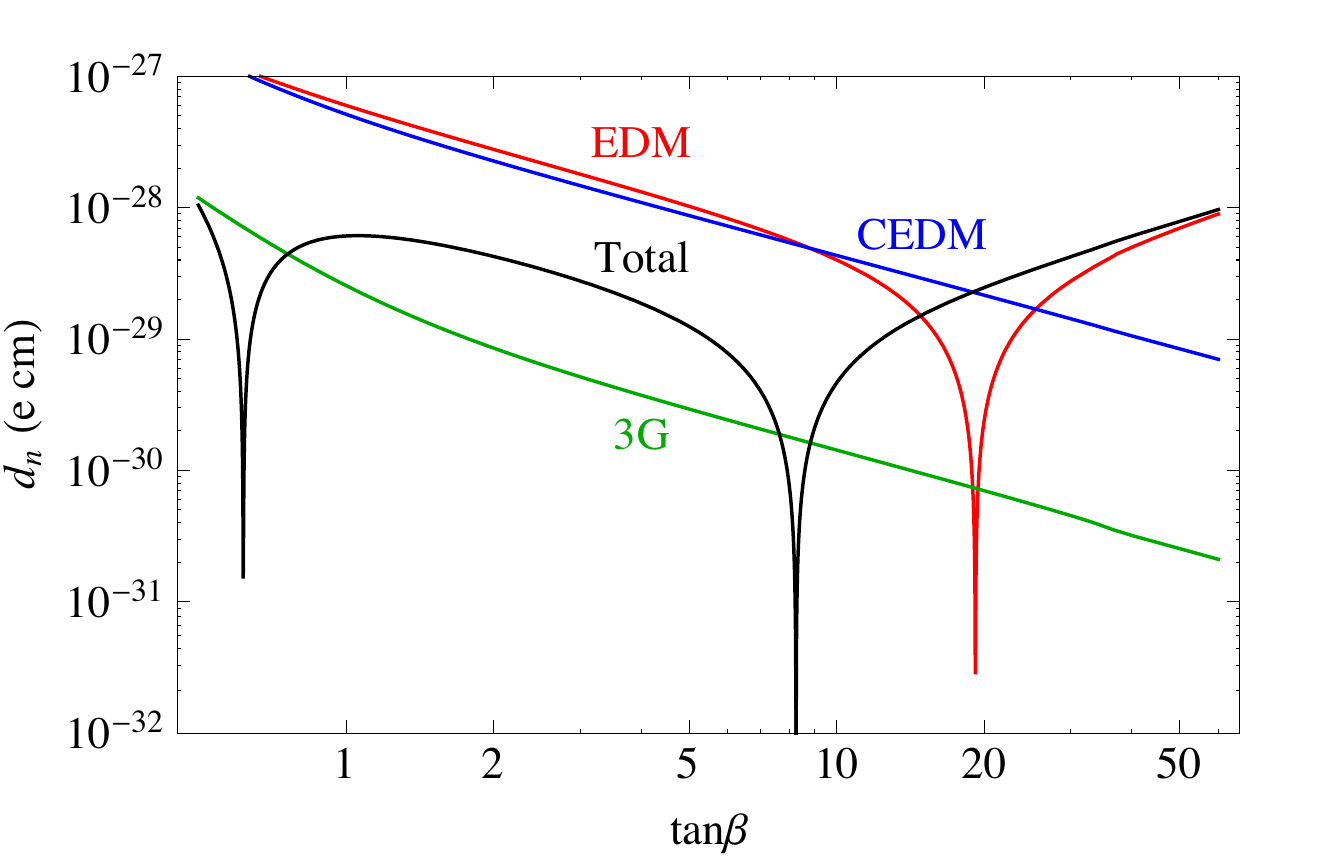}\includegraphics[width=0.5\columnwidth]{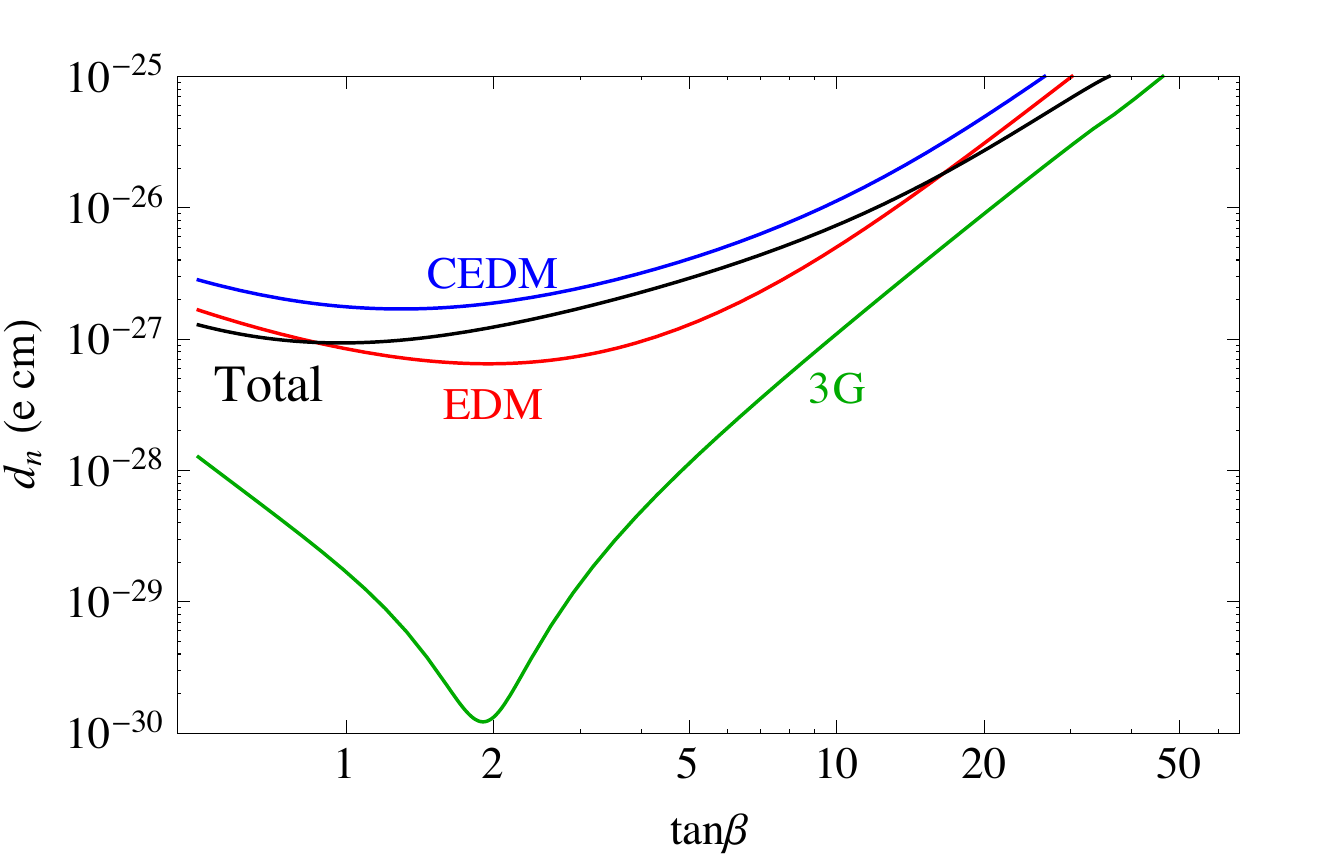}}
\caption{The anatomy of various contributions to the neutron EDM in flavor conserving 2HDMs. {\bf Left}: type-I model; {\bf Right}: type-II model.
 We plot the absolute values, so the dip in the curves implies a sign change.
The model parameters used are the same as Fig.~\ref{eEDManatomy}. 
} \label{nEDManatomy}
\end{figure}

Next, we consider the neutron EDM, whose current bound is $|d_{n}| < 2.9 \times 10^{-26} e$ cm. In Fig.~\ref{nEDManatomy}, we plot the anatomy of neutron EDM, this time in terms of the various dimension-six operator contributions. The parameters are fixed as in Fig.~\ref{eEDManatomy}, and the contributions to neutron EDM from light quark EDMs, CEDMs, and the Weinberg three-gluon operator are shown as functions of $\tan \beta$. The plot shows that in the type-II model, the quark CEDM contributions to neutron EDM are larger than that those from quark EDMs. In type-I, these two contributions are similar in size. In both cases, the effect of the Weinberg operator is smaller. Also in both types, EDM and CEDM contributions have the opposite sign, and total neutron EDM tends to be suppressed as a result. However, these statements depend on the hadronic matrix elements being close to their current best value.

\begin{figure}[h!]
\centerline{\includegraphics[width=0.5\columnwidth]{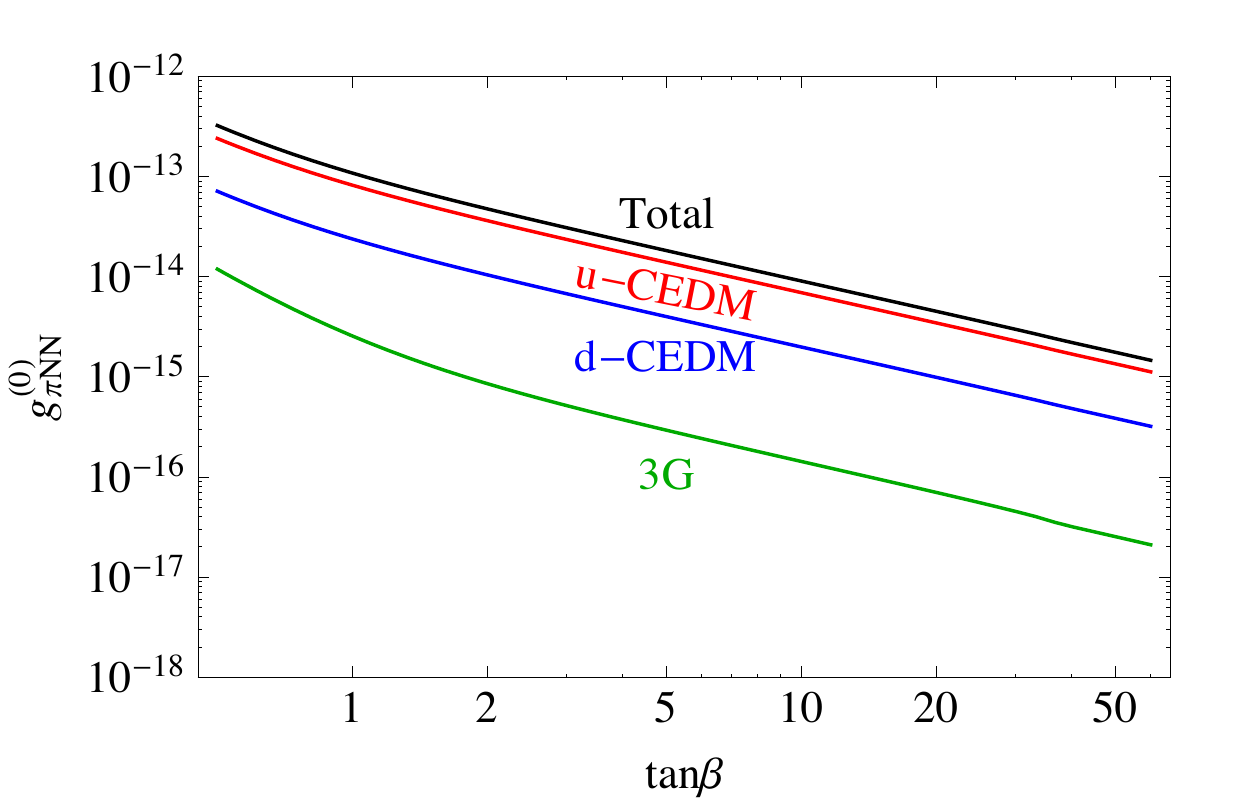}\includegraphics[width=0.5\columnwidth]{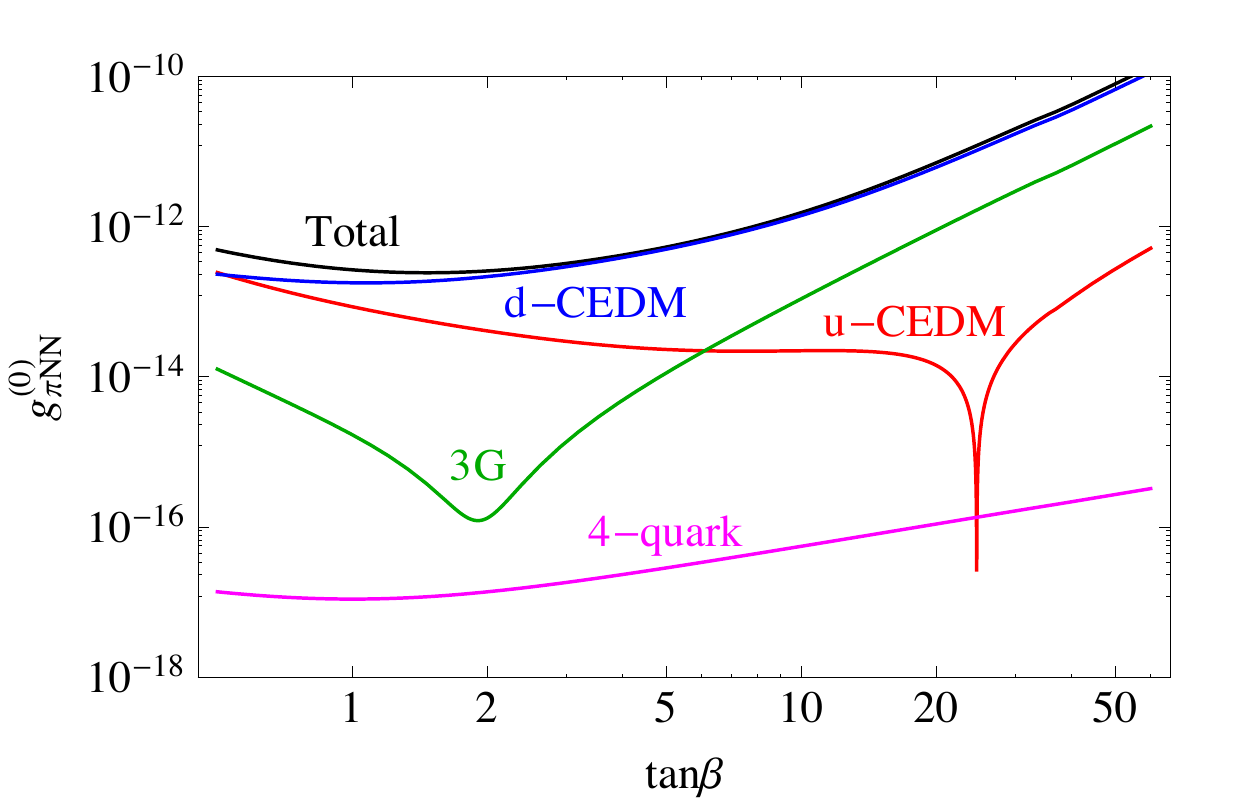}}
\caption{The anatomy of various contributions to the $\bar g_{\pi NN}^{(0)}$ for atomic EDMs in flavor conserving 2HDMs. {\bf Left}: type-I model; {\bf Right}: type-II model. 
We plot the absolute values, so the dip in the curves implies a sign change.
The model parameters used are the same as Fig.~\ref{eEDManatomy}. }\label{pinnAnatomy}
\end{figure}

The second column of Fig.~\ref{bound&uncertainty} shows the bounds in the $\sin\alpha_b$--$\tan\beta$ plane. The green regions are excluded, for three different choices of the hadronic matrix elements. Specifically, the down quark CEDM matrix element, $\tilde \zeta_n^d$, takes the values $1.63\times10^{-8}$ (solid), $0.4\times10^{-8}$ (dot-dashed), and $4.0\times10^{-8}$ (dashed). This matrix element has a large impact, because the down quark CEDM is the largest Wilson coefficient for most values of $\tan \beta$. In type-II model, in the most sensitive case with largest matrix element $\tilde \zeta_n^d = 4.0\times10^{-8}$, $\alpha_b$ is constrained to be of order 0.1 or smaller. In the least sensitive case ($\tilde \zeta_n^d = 0.4\times10^{-8}$), no part of the $\sin\alpha_b$--$\tan\beta$ plane is excluded (with $\alpha = \beta - \pi/2$). The neutron EDM constraint is quite weak in the type-I model, due to the near total cancellation of the quark EDM and CEDM contributions\footnote{This cancellation depends crucially on the relative signs of the hadronic matrix elements; see below}.

\subsection{Mercury EDM Constraint}

We now come to the $^{199}$Hg EDM limit of $|d_{Hg}| < 3.1 \times 10^{-29} e$ cm. The anatomy of the isoscalar $\pi$NN coupling $\bar{g}_{\pi NN}^{(0)}$ is shown in Fig.~\ref{pinnAnatomy}. Parameters  are chosen as in Figs.~\ref{eEDManatomy} and \ref{nEDManatomy}, and contributions from up and down CEDM, Weinberg operator are plotted as functions of $\tan \beta$. The four-quark operator involving the up and down quarks also adds to $g_{\pi NN}^{(0)}$, but this effect is negligible as shown in the type-II plot. In the type-I model, this effect is even smaller. The down quark CEDM gives the largest contribution to the $\pi$NN coupling in the type-II model, but the Weinberg operator can be important for large $\tan \beta$. In type-I, the up CEDM consistently makes up the largest bulk of $g_{\pi NN}^{(0)}$.

The parameter space excluded by the $^{199}$Hg result is plotted in the right column of Fig.~\ref{bound&uncertainty}. As for the eEDM and nEDM plots, $\alpha$ is assumed to have the SM value of $\beta - \pi/2$. The general shape of the excluded region is similar to the other two experiments, with weakest limits on $\alpha_b$ near $\tan \beta = 1$. As in the neutron case, our limits depend heavily on the value of hadronic and nuclear matrix elements. For illustration, several choices of the nuclear matrix elements $a_{0}$ and $a_{1}$ are shown. The current best values of $a_{0} = 0.01\, e$\,fm$^3$ and $a_{1} = 0.02\, e$\,fm$^3$ are represented by the solid line. In general, larger absolute values for the matrix elements imply stronger bounds, as expected, and the locations of cancellation regions are sensitive to the ratio between $a_{0}$ and $a_{1}$. In the more sensitive cases, $\alpha_b$ can be constrained to 0.1 or smaller at all values of $\tan \beta$.

\begin{figure}[t!]
\centerline{\includegraphics[width=0.5\columnwidth]{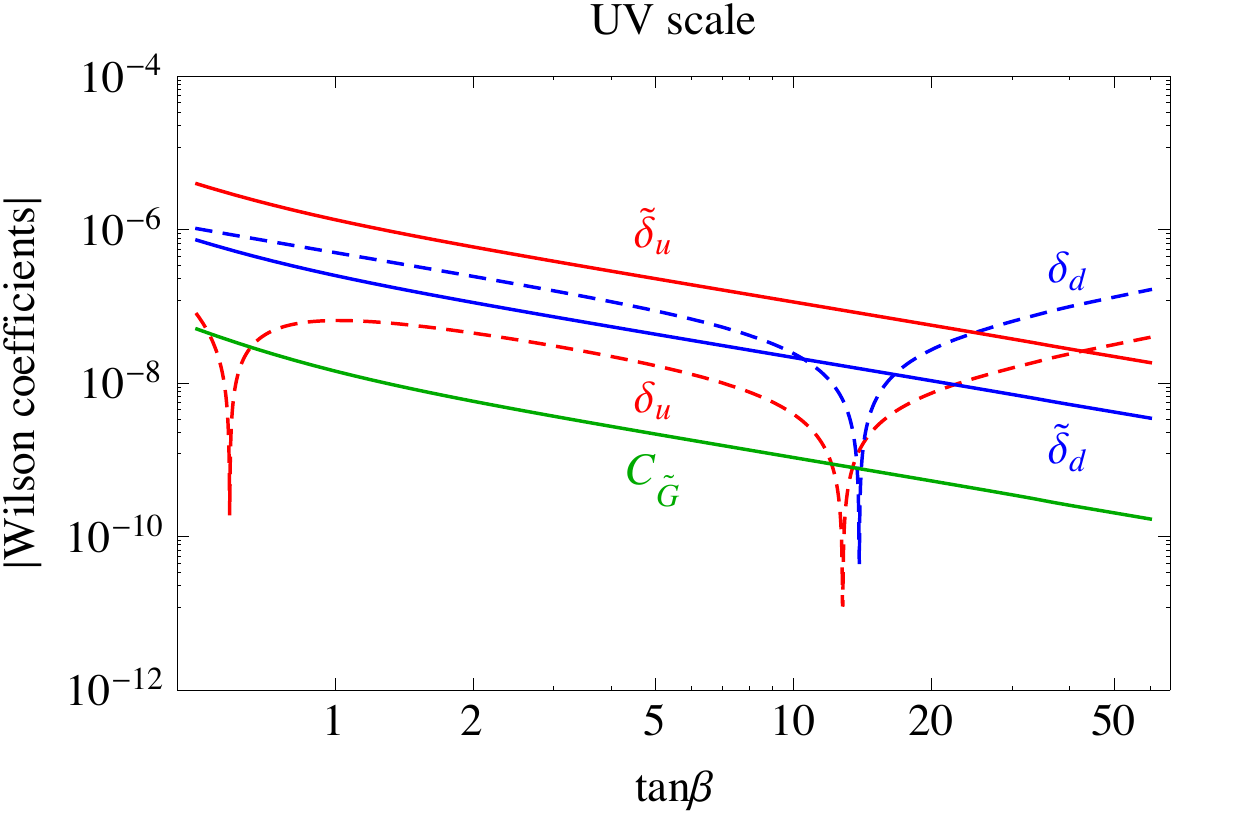}\hspace{-0.3cm}
\includegraphics[width=0.5\columnwidth]{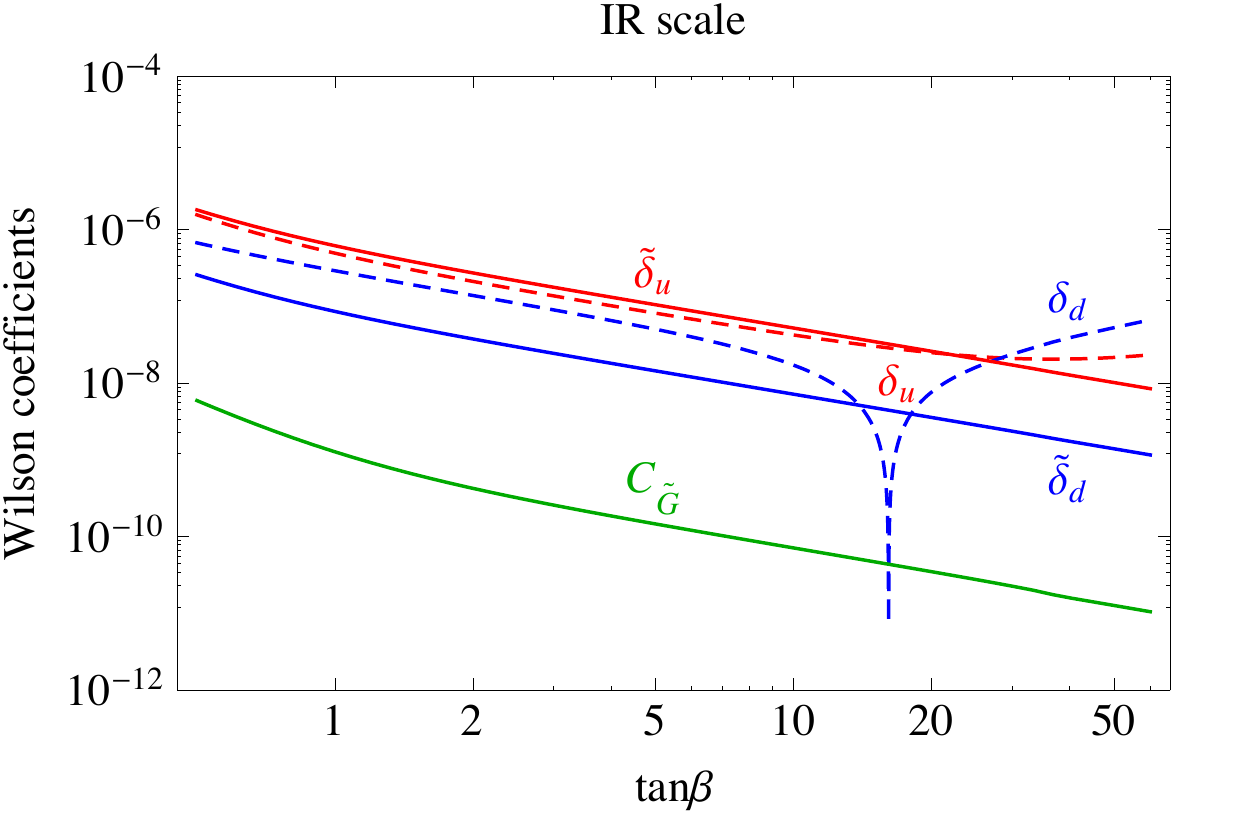}}
\centerline{\includegraphics[width=0.5\columnwidth]{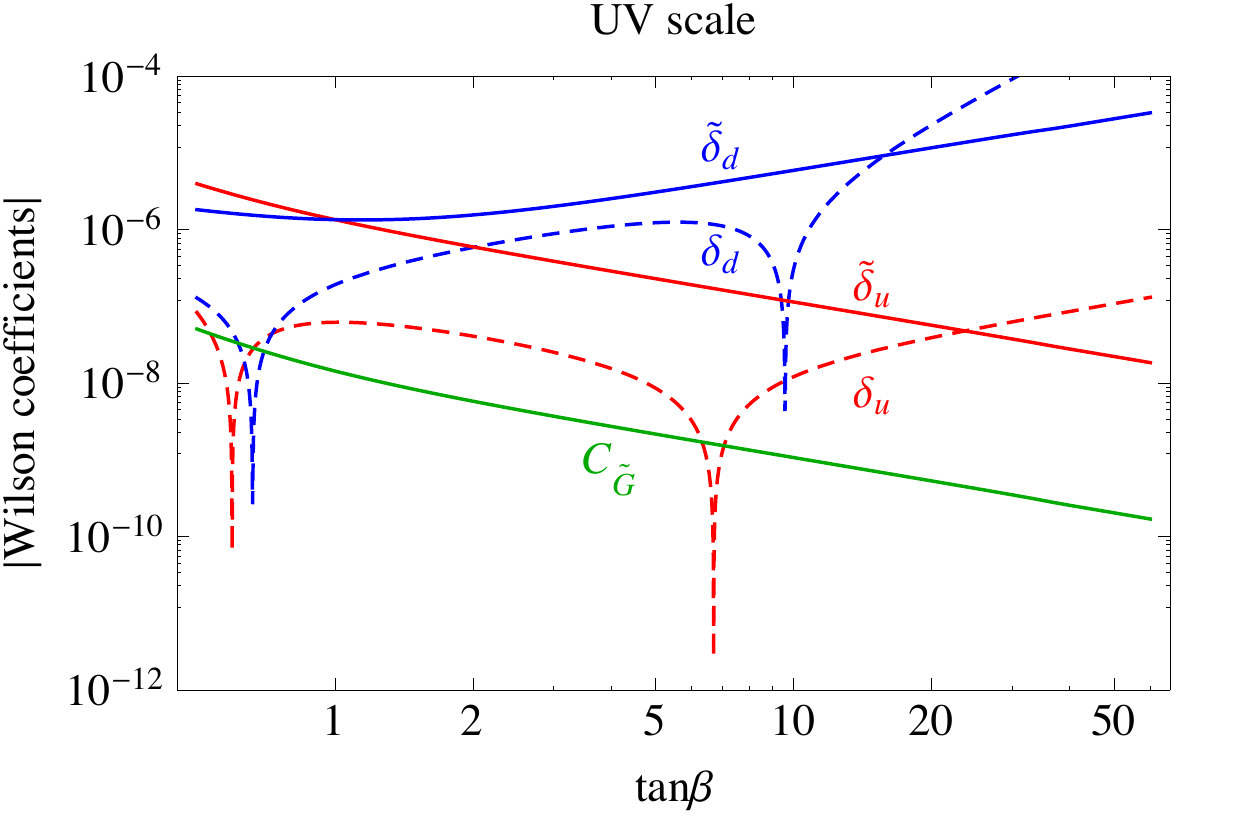}\hspace{-0.3cm}
\includegraphics[width=0.5\columnwidth]{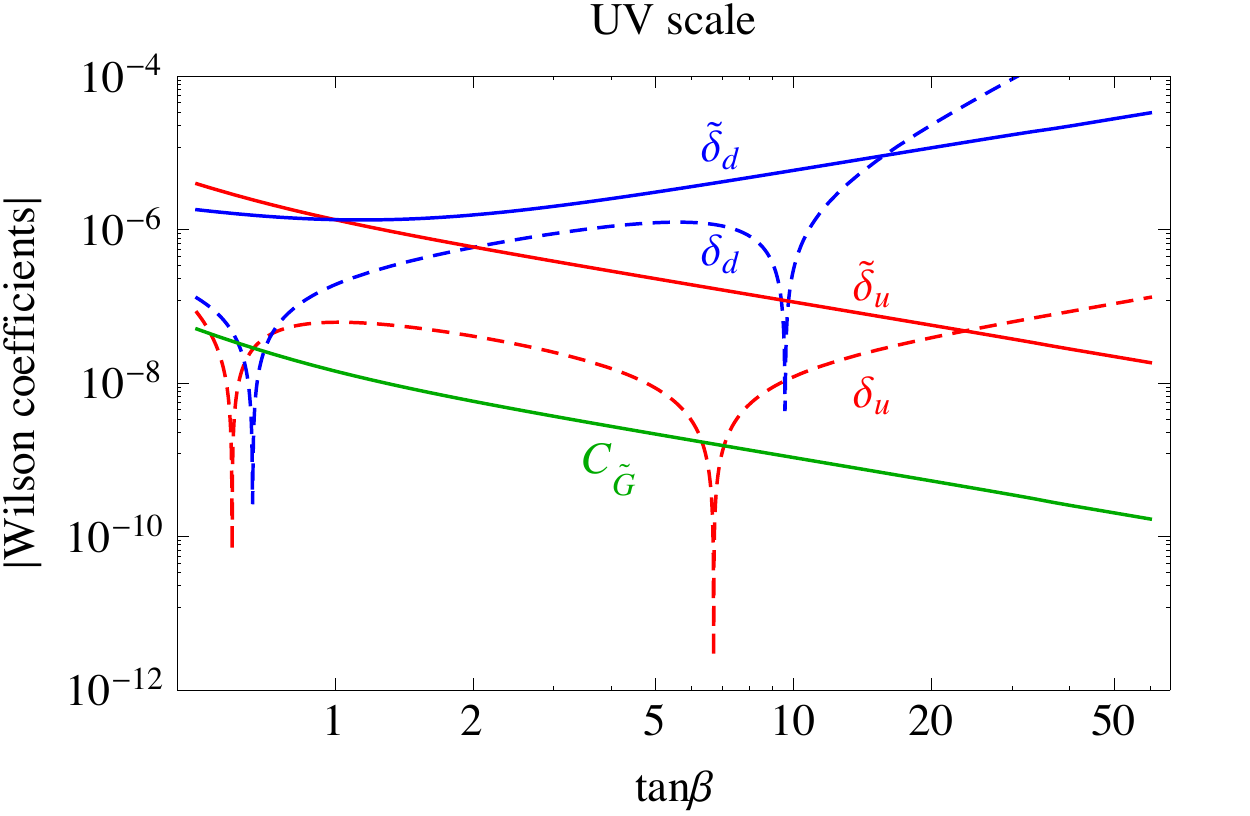}}
\caption{The Wilson coefficients at the 2HDM scale (left) and the GeV scale (right). {\bf First row}: type-I model; {\bf Second row}: type-II model.
We plot the absolute values, so the dip in the curves implies a sign change.
The differences reflect the effects of leading-order QCD corrections in the RG running. 
The model parameters used are the same as Fig.~\ref{eEDManatomy}. }\label{Fig:run}
\end{figure}

\subsection{Hadronic and nuclear uncertainties}

We have noted in the discussion of neutron and $^{199}$Hg constraints the effects of hadronic and nuclear matrix elements. Currently, calculations of these matrix elements are riddled with large uncertainties \cite{Engel:2013lsa}. For the magnitudes of the matrix elements, there is guidance from naive dimensional analysis, which takes into account the chiral structures of the operators in question. However, the precise value of matrix elements involving quark CEDMs and the Weinberg three-gluon operator are only known to about an order of magnitude, and dimensional analysis does not tell us the signs of the matrix elements. We highlight two places where these uncertainties can change our results.

\begin{itemize}
\item In Figs.~\ref{nEDManatomy} and \ref{pinnAnatomy}, we see that the Weinberg three-gluon operator is always subdominant as a contribution to the neutron and mercury EDMs. It is possible, though, that the actual matrix element may be an order of magnitude larger than the current best value. Then, the Weinberg operator would make the largest contribution to the neutron and mercury EDMs at large $\tan \beta$ in the type-II model.
\item In the left panel of Fig.~\ref{nEDManatomy}, the quark EDM and CEDM contributions to nEDM in the type-I model are shown to be nearly equal, but with opposite signs, suppressing the total neutron EDM in the type-I model. If overall sign of the CEDM matrix element is opposite to that used here,  the two effects would add constructively, making the neutron EDM limit much stronger.
\end{itemize}

In the absence of hadronic and nuclear matrix element uncertainties, improvements in neutron and diamagnetic atom searches will make them competitive with present ThO result when in constraining CPV in 2HDM. At present, however, theoretical uncertainties are significant, making it difficult to draw firm quantitative conclusions regarding the impact of the present and prospective neutron and diamagnetic EDM results.

\begin{figure}[t!]
\centerline{\hspace{0.3cm}\includegraphics[width=1.05\columnwidth]{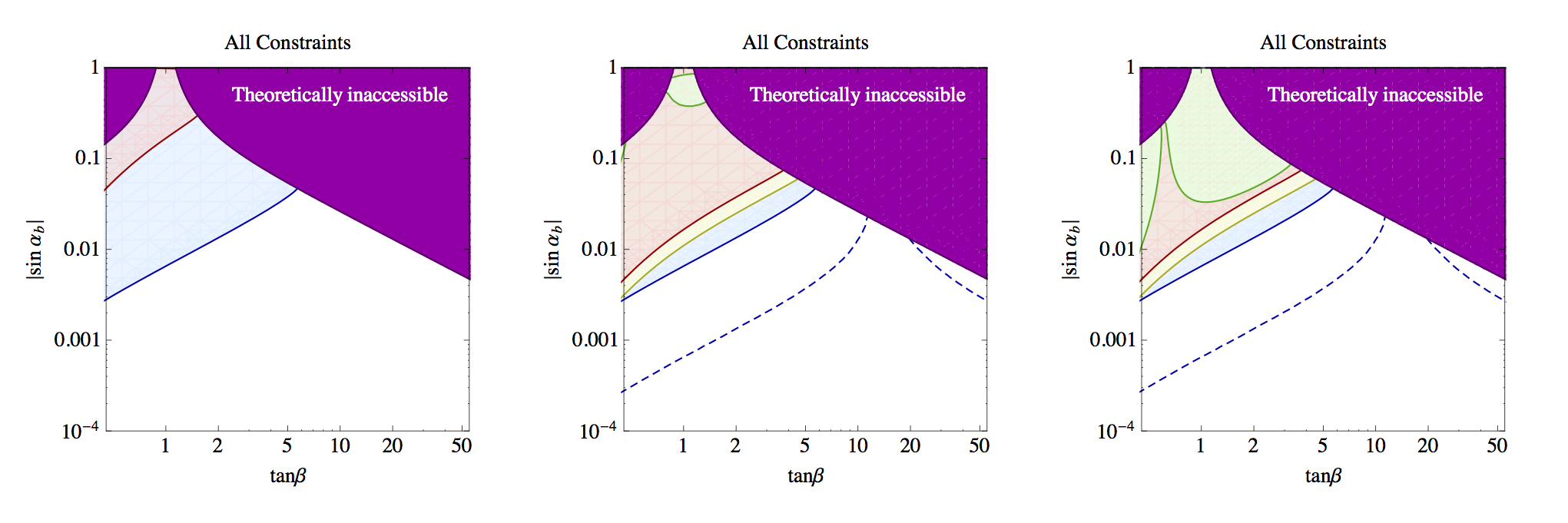}}
\centerline{\hspace{0.3cm}\includegraphics[width=1.05\columnwidth]{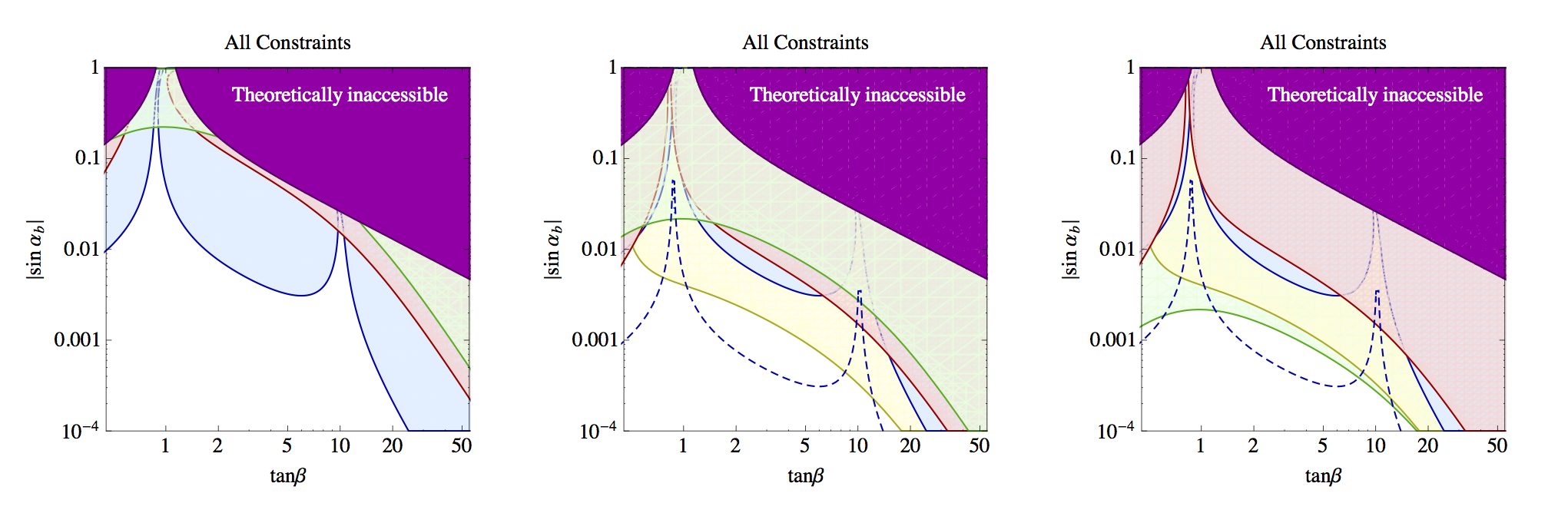}}
\caption{Current and prospective future constraints from electron EDM (blue), neutron EDM (green), Mercury EDM (red) and Radium (yellow) in flavor conserving 2HDMs. {\bf First row}: type-I model; {\bf Second row}: type-II model. 
The model parameters used are the same as Fig.~\ref{bound&uncertainty}. Central values of the hadronic and nuclear matrix elements are used. 
 {\bf Left}: Combined current limits.
{\bf Middle}: combined future limits if the Mercury and neutron EDMs are both improved by one order of magnitude. Also shown are the future constraints if electron EDM is improved by another order of magnitude (in blue dashed curves). {\bf Right}: combined future limits if the Mercury and neutron EDMs are improved by one and two orders of magnitude, respectively.} \label{Fig:Comb}
\end{figure}

\subsection{QCD Running}

Fig.~\ref{Fig:run} illustrates the differences in the Wilson coefficients between the UV (weak) scale and the IR (hadronic) scale. 
For the type-I model (top panels), there is no dramatic difference between magnitude of the coefficients other than a slight enhancement in quark EDMs and CEDMs. For the type-II model (bottom panels), however, there is a significant difference in the $\tan \beta$ dependence and the relative magnitudes of the coefficients. Part of this difference is explained by diagrams involving b quarks, which is not yet integrated out at the UV scale. Specifically, much of the growing behaviors at large $\tan \beta$ for the d-quark CEDM and the Weinberg operator are accounted for by simply adding the b-quark diagrams. However, the growth of the u-quark EDM and u-quark CEDM at large $\tan \beta$, and the similarity of EDM and CEDM for each quark at all values of $\tan \beta$ are mainly due to the operator mixing in QCD. For this reason, we observe that QCD running must be taken into account when making quantitative claims about hadronic sources of CP violation.

\subsection{Combined EDM Constraints: Present and Future}

We summarize the combined EDM constraints of the electron (blue region), neutron (green), $^{199}$Hg (red) and $^{225}$Ra (yellow) in Fig.~\ref{Fig:Comb}. For the neutron and atomic EDMs, we use the central values for the hadronic and nuclear matrix elements. The first column shows the present constraints. We find that the bound on electron EDM from ACME experiment is presently by far the strongest, except for the cancellation region, which is closed by the mercury and neutron EDM bounds. The current constraints are roughly, $\alpha_b, \alpha_c\lesssim0.1$ for all $\tan\beta$ in type-II 2HDM. For the type-I model, the constraints are $\alpha_b\lesssim0.1$, while $\alpha_c$ can still be order one for $\tan\beta\gtrsim5$. Therefore, it could be easier to search for CPV effects related to the heavy scalars if they are discovered in the future collider experiments.

The second column of Fig.~\ref{Fig:Comb} shows the future constraints if the neutron and Mercury EDM experiments improve the current sensitivities by a factor of 10. 
We have also shown the future constraints in blue dashed curves if electron EDM is improved by another order of magnitude.
The last column shows the situation when the neutron EDM limit is improved by a factor 100, which is the goal for the experiment planned for the Fundamental Neutron Physics Beamline at the Oak Ridge National Laboratory Spallation Neutron Source. We find for type-II 2HDM, the future neutron EDM experiments can improve the current limit on the CPV angles $\alpha_b, \alpha_c$ by one order of magnitude. The radium EDM also has the prospect to give a comparable limit.

Finally, we comment on the effects of changing the masses of the heavy scalars. While we have only presented the results for the case when the extra scalars have masses of order $\sim 400$ GeV, we have performed the same analyses with different masses of up to $\sim 500$ GeV. The constraints on the CPV angle $\alpha_b$ become slightly weaker with heavier new scalars, but we find that there is no qualitative difference to our conclusions.

%

\section{Summary}
\label{sec:summary}
The nature of CPV beyond the Standard Model remains a question at the forefront of fundamental physics. The cosmic matter-antimatter asymmetry strongly implies that such BSM CPV should exist, but the associated mass scale and dynamics remain unknown. With the observation of the 125 GeV boson at the LHC, it is particularly interesting to ask whether the scalar sector of the larger framework containing the SM admits new sources of CPV and, if so, whether their effects are experimentally accessible. In this study, we have explored this question in the context of flavor conserving 2HDMs, allowing for a new source of CPV in the scalar potential. The present constraints on this type of CPV are generally weaker than for scenarios where the BSM directly enters the couplings to SM fermions, as the associated contributions to electric dipole moments generically first appear at two-loop order. In this context, we find that present EDM limits are complementary to scalar sector constraints from LHC results, as the latter generally constrain the CP-conserving sector of the type-I and type-II models, whereas EDMs probe the CPV parameter space. Moreover, despite the additional loop suppression, the present ThO, $^{199}$Hg, and neutron EDM search constraints are quite severe, limiting $|\sin\alpha_b|$ to $\sim 0.01$ or smaller for most values of $\tan \beta$.

The next generation of EDM searches could extend the present reach by an order of magnitude or more and could allow one to distinguish between the type-I and type-II models. In particular, a non-zero neutron or diamagnetic atom EDM result would likely point to the type-II model, as even the present ThO limit precludes an observable effect in the type-I scenario given the planned sensitivity of the neutron and diamagnetic atom searches. Furthermore, it appears that a combination of searches using different systems would be needed to achieve a comprehensive probe of the relevant parameter space in the type-II model. We emphasize, however, that these expectations are somewhat provisional, given the present substantial uncertainties associated with computations of the hadronic and nuclear matrix elements that we have quantified in this study. Achieving more robust computations would be particularly welcome, especially given the role of neutron and diamagnetic atom EDM searches in probing the type-II flavor conserving 2HDM. 

\bigskip
\noindent \noindent{\it  Acknowledgements} 
We thank J. DeVries and W. Dekens for helpful discussion of the effective operator anomalous dimension matrix. 
The work is partially supported by the Gordon and Betty Moore Foundation through Grant No.\,776 to the Caltech Moore Center for Theoretical Cosmology and Physics and by DOE Grant DE-FG02-92ER40701 (YZ) and DE-SC0011095 (SI and MJRM). The work of YZ is also supported by a DOE Early Career Award under Grant No. DE-SC0010255.

\appendix

\section{Wilson coefficients of P and T-odd operators at the 2HDM scale}\label{alledms}

In this appendix, we give the results of all the Wilson coefficients by integrating out the heavy particles at the 2HDM scale, $\Lambda \sim M_Z$. At this scale, the bottom quark is still light, and we discuss the matching conditions at the $m_b$ scale in subsection~\ref{running}. The total contributions to the Weinberg, CEDM and EDM operators are Eq.~(\ref{A1}), (\ref{A4}) and (\ref{A15}) respectively.

%

\begin{figure*}[h!]
  \centerline{\includegraphics[width=1.0\columnwidth]{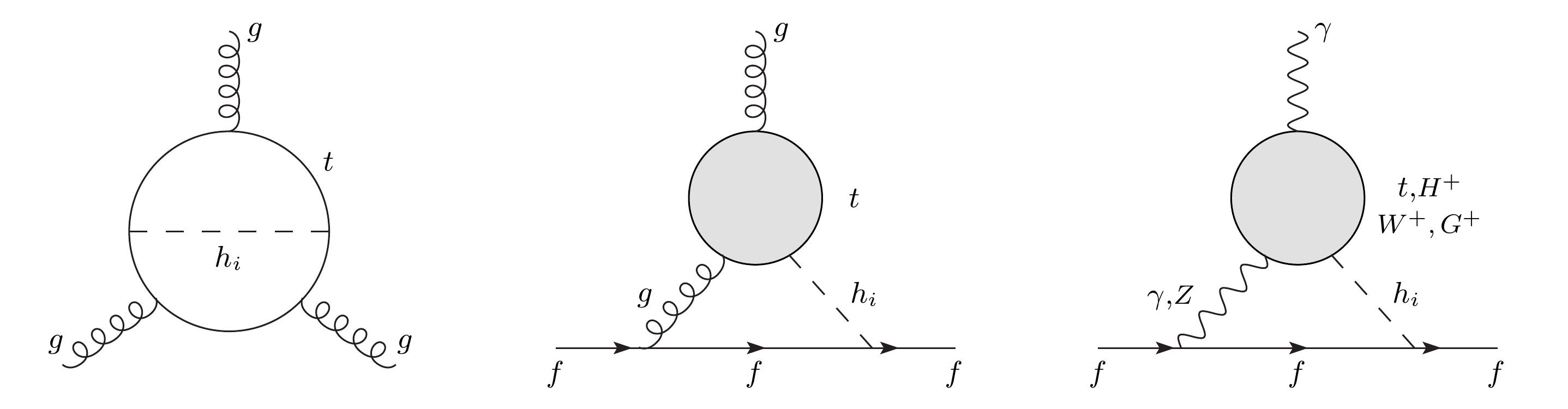}}
\caption{{\bf Left}: two-loop contribution to the Weinberg operator with CPV neutral Higgs mixings.
{\bf Middle}: quark CEDM from Barr-Zee type diagrams with $g h_i$ exchange and CPV neutral Higgs mixings. The conjugate diagrams are not shown.
{\bf Right}: quark or lepton EDM from Barr-Zee type diagrams with $\gamma h_i$ or $Z h_i$ exchange and CPV neutral Higgs mixings.
}\label{twoloop}
\end{figure*}

\subsection{Two-loop Weinberg Operator}
From Ref.~\cite{Weinberg:1989dx}, the contribution to the $d=6$ Weinberg operator arises from the top loop, as shown in the left panel of Fig.~\ref{twoloop}, which gives
\begin{eqnarray}\label{A1}
C_{\tilde{G}} (\Lambda) \equiv (C_{\tilde{G}})_t 
= - \frac{g_s^2}{3} \frac{1}{128 \pi^4} \sum_{i=1}^3 h_0(m_t/m_{h_i}) c_{t,i} \tilde c_{t,i} \ ,
\end{eqnarray}
where the function $h_0(x)$ can be found in the Appendix~\ref{loopf}.

\subsection{Two-loop Barr-Zee type contributions to CEDMs}

For light fermions, the dominant contributions to their EDMs and CEDMs come from the two-loop Barr-Zee type diagrams~\cite{Barr:1990vd}, as shown in the middle and right panels of Fig.~\ref{twoloop}.

For the CEDM, the top quark in the upper (shaded) loop is first integrated out to obtain the $h_iGG$ or $h_iG\tilde G$ operators, which then contribute to the CEDM operators~\cite{Gunion:1990iv},
\begin{eqnarray}\label{A4}
\tilde{\delta}_q(\Lambda) \equiv \left(\tilde{\delta}_q \right)^{hgg}_{t} = - g_s^2 \frac{1}{128\pi^4} \sum_{i=1}^3 \left[ f(z^i_{t}) c_{t,i} \tilde c_{q,i} + g(z^i_{t}) \tilde c_{t,i} c_{q,i} \right] \ ,
\end{eqnarray}
where $q=u,d,b$, and $z^i_{t}= m_{f_1}^2/m_{h_i}^2$. The two-loop functions $f(x), g(x)$ can be found in the Appendix~\ref{loopf}.

\subsection{Two-loop Barr-Zee type contributions to EDMs: diagrams with $H^0\gamma \gamma$ and $H^0 Z \gamma$}

The corresponding Barr-Zee type EDMs for light fermions are obtained with operators $h_i FF$ or $h_i F\tilde F$ from the upper (shaded) loop. See the right panel of Fig.~\ref{twoloop}. The contribution from top quark is
\begin{eqnarray}\label{A6}
\left(\delta_f \right)^{h\gamma\gamma}_{t} = - N_c Q_f Q_{t}^2 e^2 \frac{1}{64\pi^4} \sum_{i=1}^3 \left[ f(z^i_{t}) c_{t,i} \tilde c_{f,i} + g(z^i_{t}) \tilde c_{t,i} c_{f,i} \right] \ .
\end{eqnarray}
Here the external fermions relevant for our calculations are $f=u,d,e$. The analog contribution is to replace the photon propagator with that of the $Z$-boson. It is worth noting that only the vector current of the $Z\bar ff$ coupling enters in the final EDM, which is 
\begin{eqnarray}\label{A7}
\left(\delta_f \right)^{hZ\gamma}_{t} = - N_c Q_{f_1} g_{Z\bar f f}^V g_{Z\bar tt}^V \frac{1}{64\pi^4} \sum_{i=1}^3 \left[ \tilde f(z^i_{t}, m_{t}^2/M_Z^2) c_{t,i} \tilde c_{f,i} + \tilde g(z^i_{t}, m_{t}^2/M_Z^2) \tilde c_{t,i} c_{f,i} \right] \ ,
\end{eqnarray}
with 
$g_{f\bar f Z}^V = g(T_3^f - 2 Q^f \sin^2\theta_W)/({2\cos\theta_W}) $.
The loop function $\tilde f(z,x)$ and $\tilde g(z,x)$ can be found in the Appendix~\ref{loopf}. The corresponding bottom quark loop contribution is properly taken into account in sec.~\ref{running}.

\bigskip

In the right panel of Fig.~\ref{twoloop}, the particles in the upper (shaded) loop can also be the $W$-boson and its Goldstone boson. The gauge invariant contributions have been obtained in~\cite{Chang:1990sf, Abe:2013qla},
\begin{eqnarray}\label{gammaW}
\left(\delta_f \right)^{h\gamma\gamma}_W &=& Q_f e^2 \frac{1}{256\pi^4} \sum_{i=1}^3 \left[ \left( 6 + \frac{1}{z^i_w} \right) f(z^i_w) + \left( 10 - \frac{1}{z^i_w} \right) g(z^i_w) \right] a_i \tilde c_{f,i}  \ , \label{A8} \\
\left(\delta_f \right)^{hZ\gamma}_W &=& g_{Z\bar f f}^V g_{ZWW} \frac{1}{256\pi^4} \sum_{i=1}^3 \left[ \left(6 -\sec^2\theta_W + \frac{2-\sec^2\theta_W}{2z^i_w} \right)\tilde f(z^i_w, \cos^2\theta_W) \right.\nonumber \\
&&\hspace{2cm}+ \left. \left( 10- 3\sec^2\theta_W - \frac{2-\sec^2\theta_W}{2z^i_w}\right)\tilde g(z^i_w, \cos^2\theta_W) \right] a_i \tilde c_{f_i} \ , \label{A9}
\end{eqnarray}
where $z^i_w = M_W^2/m_{h_i}^2$ and $g_{WWZ}/e = \cot\theta_W$.

\bigskip

Similarly, the physical charged scalar can also run in the loop of Fig.~\ref{twoloop}. This is similar to the squark contribution discussed in the supersymmetric framework~\cite{Chang:1998uc}. With the couplings defined in Eq.~(\ref{lamtil}), its contributions to EDM are
\begin{eqnarray}\label{gammaH+}
\left(\delta_f \right)^{h\gamma\gamma}_{H^+} &=& Q_f e^2 \frac{1}{256\pi^4} \left(\frac{v}{m_{H^+}}\right)^2 \sum_{i=1}^3\left[ f(z^i_H) - g(z^i_H)\right] \bar \lambda_i \tilde c_{f,i} \ ,\label{A10} \\
\left(\delta_f \right)^{hZ\gamma}_{H^+} &=& g_{Z\bar f f}^V g_{ZH^+ H^-} \frac{1}{256\pi^4} \left( \frac{v}{m_{H^+}}\right)^2 \sum_{i=1}^3 \left[ \tilde f(z^i_H, m_{H^+}^2/M_Z^2) - \tilde g(z^i_H, m_{H^+}^2/M_Z^2)\right] \bar \lambda_i \tilde c_{f,i} \ , \label{A11}
\end{eqnarray}
with $z^i_H = m_{H^+}^2/m_{h_1}^2$ and $g_{ZH^+H^-}/e = \cot\theta_W(1-\tan^2\theta_W)/2$.

\subsection{Two-loop Barr-Zee type contributions to EDMs: diagrams with $H^\pm W^\mp \gamma$}

\begin{figure*}[h!]
  \centerline{\includegraphics[width=.8\columnwidth]{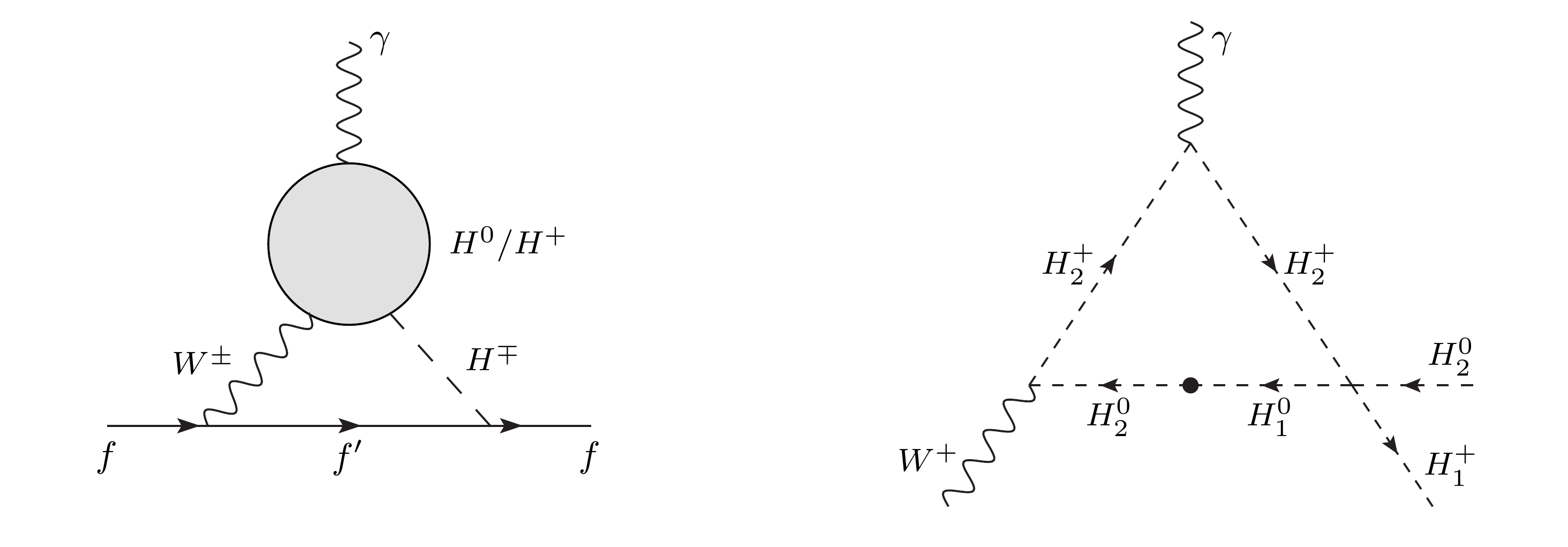}}
\caption{{\bf Left}: quark or lepton EDM from $W^\pm H^\mp$ exchange and CPV Higgs interactions.
{\bf Right}: a scalar loop contribution to $\phi_1^\dagger \frac{\sigma^a}{2} W_{\mu\nu}^a \phi_2 B^{\mu\nu}$ effective operator, which then contributes to EDM as the upper loop of the left panel.}
  \label{34}
\end{figure*}

The left panel of Fig.~\ref{34} represents the contribution where the upper loop yields an $H^\pm W^\mp \gamma$ operator. This contribution has not been included in the 2HDM calculations until very recently~\cite{Abe:2013qla}  (see~\cite{Chang:1999zw} for the counterpart in supersymmetric models).

Here we would like to stress that it arises from the only source of CP violation in the Higgs potential.
In the effective theory language, the possible gauge invariant operators for the upper (shaded) loop include
\begin{eqnarray}\label{operators}
C_{ij} \phi_i^\dagger \frac{\sigma^a}{2} W_{\mu\nu}^a \phi_j B^{\mu\nu}, \ \ \ \tilde C_{ij} \phi_i^\dagger \frac{\sigma^a}{2} W_{\mu\nu}^a \phi_j \tilde B^{\mu\nu} \ ,
\end{eqnarray}
where $C_{ij}, \tilde C_{ij}$ are the Wilson coefficients. Because of the CP properties, fermion EDMs are proportional to the imaginary part of $C_{ij}$ or the real part of $\tilde C_{ij}$. Here we argue that in the flavor conserving 2HDMs discussed in this work, only the scalar loop could contribute to $C_{12}$ and eventually to EDMs. A representative diagram is shown in the right panel of Fig.~\ref{34}. It is proportional to
\begin{eqnarray}
{\rm Im} (\lambda_5 m_{12}^{2*} v_1^* v_2) = - \left| \lambda_5 m_{12}^{2} v_1 v_2 \right| \sin \delta_2 \ .
\end{eqnarray}
Using the relation in Eq.~(\ref{eq:invar2}), the above quantity is indeed related to the unique CPV source in the model. 

The fermionic loops do not contribute because the physical charge Higgs and quark couplings have the structure proportional to the corresponding CKM element. As a result,
the coefficients $C_{ij}$ are purely real and $\tilde C_{ij}$ are purely imaginary. They contribute to magnetic dipole moments instead of EDMs.

The gauge invariant contributions to EDM from this class of diagrams have been calculated recently in~\cite{Abe:2013qla},
\begin{eqnarray}\label{A14}
\left(\delta_f \right)^{HW\gamma}_H &=& \frac{1}{512\pi^4} s_f \sum_i \left[ \frac{e^2}{2\sin^2\theta_W} \mathcal{I}_4(m_{h_i}^2, m_{H^+}^2) a_i \tilde c_{f,i} - \mathcal{I}_5(m_{h_i}^2, m_{H^+}^2) \bar \lambda_i \tilde c_{f,i} \right] \ ,
\end{eqnarray}
where the functions $\mathcal{I}_{4,5}(m_1^2, m_2^2)$ are given in the Appendix~\ref{loopf}. The coefficient $s_f=-1$ for up-type quarks, and  $s_f=+1$ for down-type quarks and charged leptons. 

\bigskip
To summarize, the total contribution to fermion EDM is the sum of Eqs~(\ref{A6},\ref{A7},\ref{A8},\ref{A9},\ref{A10},\ref{A11},\ref{A14}),
\begin{eqnarray}\label{A15}
\delta_f (\Lambda) \equiv \left(\delta_f \right)^{h\gamma\gamma}_{t} + \left(\delta_f \right)^{hZ\gamma}_{t} + \left(\delta_f \right)^{h\gamma\gamma}_{W} + \left(\delta_f \right)^{hZ\gamma}_{W} +\left(\delta_f \right)^{h\gamma\gamma}_{H^+} + \left(\delta_f \right)^{hZ\gamma}_{H^+} + \left(\delta_f \right)^{HW\gamma}_H \ .
\end{eqnarray}

\section{Loop functions} 
\label{loopf}

In this appendix, we collect all the loop functions used in previous sections.
\begin{eqnarray}
h_0(z) &=& \frac{z^4}{2} \int_0^1 dx \int_0^1 dy \frac{x^3 y^3 (1-x)}{(z^2x(1-xy)+(1-y)(1-x))^2} \ , \\
f(z) &=& \frac{z}{2} \int_0^1 dx \frac{1-2x(1-x)}{x(1-x)-z} \log\frac{x(1-x)}{z} \ , \\
g(z) &=& \frac{z}{2} \int_0^1 dx \frac{1}{x(1-x)-z} \log\frac{x(1-x)}{z} \ , \\
h(z) &=& \frac{z}{2} \int_0^1 dx \frac{1}{z-x(1-x)} \left( 1+ \frac{z}{z-x(1-x)} \log\frac{x(1-x)}{z} \right) \ , \\
\tilde f(x,y) &=& \frac{yf(x)}{y-x} + \frac{x f(y)}{x-y}  \ , \\
\tilde g(x,y) &=& \frac{yg(x)}{y-x} + \frac{x g(y)}{x-y} \ , \\
\mathcal{I}_{4,5}(m_1^2, m_2^2) &=& \frac{M_W^2}{m_{H^+}^2 - M_W^2} \left(I_{4,5}(M_W^2, m_1^2)-I_{4,5}(m_2^2, m_1^2)\right) \ , \\
I_4(m_1^2,m_2^2) &=& \int_0^1 dz (1-z)^2 \left( z-4 + z\frac{m_{H^+}^2 -m_2^2}{M_W^2} \right)  \nonumber \\
&&\hspace{0.6cm} \times \frac{m_1^2}{M_W^2(1-z)+m_2^2 z - m_1^2z(1-z)} \log\frac{M_W^2(1-z)+m_2^2z}{m_1^2z(1-z)} \ , \\
I_5(m_1^2,m_2^2) &=& \int_0^1 dz \frac{m_1^2 z(1-z)^2}{M_W^2(1-z)+m_2^2 z - m_1^2z(1-z)} \log\frac{M_W^2(1-z)+m_2^2z}{m_1^2z(1-z)} \ .
\end{eqnarray}

\end{document}